\documentclass[aps,prb,preprintnumbers,superscriptaddress,amsmath,floatfix,twocolumn]{revtex4-2}
\usepackage{listings}
\usepackage{verbatim}
\usepackage{graphicx} 
\usepackage{subcaption}

\usepackage{graphicx}
\usepackage{dcolumn}
\usepackage{bm}
\usepackage{soul}
\usepackage{color}
\usepackage{xcolor}
\usepackage{amsmath}
\usepackage{amssymb}
\usepackage{float}
\usepackage{natmove}
\usepackage{multirow}
\usepackage{setspace}
\usepackage{array}
\usepackage{booktabs}
\usepackage{rotating}
\usepackage{ulem}
\usepackage[colorlinks=true,linkcolor=black,citecolor=black,urlcolor=black]{hyperref}
\usepackage{physics}
\usepackage{mathrsfs}
\usepackage{cprotect}
\usepackage{mathtools}
\usepackage[capitalise]{cleveref}
\usepackage[version=4]{mhchem} 
\crefname{figure}{Fig.}{Figs.}
\Crefname{figure}{Fig.}{Figs.}
\crefname{table}{Table}{Tables}
\Crefname{table}{Table}{Tables}
\crefname{equation}{Eq.}{Eqs.}
\Crefname{equation}{Eq.}{Eqs.}
\crefname{section}{Section}{Sections}
\Crefname{section}{Section}{Sections}
\usepackage[acronym,nomain,nopostdot,nogroupskip,nostyles]{glossaries}
\newacronym{bse}{BSE}{Bethe-Salpeter equation}
\newacronym{mbpt}{MBPT}{many-body perturbation theory}
\newacronym{qp}{QP}{quasiparticle}
\newacronym{dft}{DFT}{density-functional theory}
\newacronym{dfpt}{DFPT}{density-functional perturbation theory}
\newacronym{ks}{KS}{Kohn-Sham}
\newacronym{kc}{KC}{Koopmans-compliant}
\newacronym{pwl}{PWL}{piecewise-linear}
\newacronym{qki}{qKI}{quadratic Koopmans-integral}
\newacronym{mlwf}{MLWF}{maximally localized Wannier function}
\newacronym{rpa}{RPA}{random-phase approximation}
\newacronym{lda}{LDA}{local-density approximation}
\newacronym[sort=QSGW]{qsgw}{QSG$\widetilde{W}$}{quasiparticle self-consistent
  $GW$ with vertex corrections in the screened Coulomb interaction}

\newacronym{ebe}{EBE}{exciton binding energy}
\newacronym{bz}{BZ}{Brillouin zone}
\newacronym{zpr}{ZPR}{zero-point renormalization}
\newacronym{soc}{SOC}{spin-orbit coupling}
\newacronym{nc}{NC}{non-collinear}

\newacronym{eels}{EELS}{electron energy-loss spectroscopy}
\newacronym{nrixs}{NRIXS}{non-resonant inelastic X-ray scattering}

\newacronym{mare}{MARE}{mean absolute relative error}
\newacronym{si}{SI}{Supplementary Information}

\definecolor{Grey}{rgb}{0.50,0.50,0.50}
\definecolor{Blu}{rgb}{0.00,0.00,1.00}
\definecolor{Red}{rgb}{1.00,0.00,0.00}
\definecolor{Green}{rgb}{0.1,0.9,0.0}
\definecolor{Magenta}{rgb}{0.60,0.00,0.60}
\definecolor{BluBondi}{rgb}{0.00,0.58,0.71}
\definecolor{Orange}{rgb}{0.95,0.46,0.17}
\definecolor{Red}{rgb}{1.00,0.00,0.00}

\newcommand{\editor}[2]{%
  \expandafter\newcommand\csname #1note\endcsname[1]{%
    \textcolor{#2}{{\it (\textbf{#1:} ##1)}}}%
  \expandafter\newcommand\csname #1\endcsname[1]{%
    \textcolor{#2}{##1}}%
  \expandafter\newcommand\csname #1cancel\endcsname[1]{%
    \textcolor{#2}{\sout{##1}}}%
  \expandafter\newcommand\csname #1change\endcsname[2]{%
    \textcolor{#2}{\sout{##1} ##2}}%
  \newenvironment{#1text}{\color{#2}}{\color{black}}
}

\editor{MB}{Green}     
\editor{NC}{Magenta}   
\editor{EL}{BluBondi}   
\editor{NM}{BluBondi}  

\newcommand{\drop}[1]{}

\renewcommand{\emph}{\textit}

\def\QE{\textsc{Quantum ESPRESSO}}   
\usepackage{adjustbox}
\begin{document}
\title{Koopmans spectral functionals for excitons in materials}
%
\author{Miki Bonacci}
\email[corresponding author: ]{miki.bonacci@nano.cnr.it}
\affiliation{Istituto Nanoscienze -- CNR, S3, Via G. Campi 213/A, Modena, Italy}
\affiliation{PSI Center for Scientific Computing, Theory and Data, 5232 Villigen PSI, Switzerland}
\affiliation{
National Centre for Computational Design and Discovery of Novel Materials (MARVEL), 5232 Villigen PSI, Switzerland
}
\author{Edward Linscott}
\affiliation{PSI Center for Scientific Computing, Theory and Data, 5232 Villigen PSI, Switzerland}
\affiliation{
National Centre for Computational Design and Discovery of Novel Materials (MARVEL), 5232 Villigen PSI, Switzerland
}
\author{Julian Geiger}
\affiliation{PSI Center for Scientific Computing, Theory and Data, 5232 Villigen PSI, Switzerland}
\affiliation{
National Centre for Computational Design and Discovery of Novel Materials (MARVEL), 5232 Villigen PSI, Switzerland
}
\author{Nicola Marzari}
\affiliation{PSI Center for Scientific Computing, Theory and Data, 5232 Villigen PSI, Switzerland}
\affiliation{
National Centre for Computational Design and Discovery of Novel Materials (MARVEL), 5232 Villigen PSI, Switzerland
}
\affiliation{Theory and Simulation of Materials (THEOS), \'Ecole Polytechnique F\'ed\'erale de Lausanne, CH-1015 Lausanne, Switzerland}
\affiliation{Theory of Condensed Matter, Cavendish Laboratory, University of Cambridge, Cambridge CB3 0US, United Kingdom}
\author{Nicola Colonna}
\affiliation{PSI Center for Scientific Computing, Theory and Data, 5232 Villigen PSI, Switzerland}
\affiliation{
National Centre for Computational Design and Discovery of Novel Materials (MARVEL), 5232 Villigen PSI, Switzerland
}
\begin{abstract}
Accurate first-principles characterization of excitons typically relies on the \gls{bse} combined with quasiparticle energies from computationally demanding $GW$ calculations. Here, we replace the $GW$ step with quasiparticle energies from Koopmans-compliant spectral functionals, an approach we call BSE@qKI, and implement it in an automated, reproducible AiiDA workflow. For a set of benchmark semiconductors, BSE@qKI reproduces experimental band gaps and first exciton energies with mean absolute relative errors of 13\% and 4\%, respectively, comparable to BSE@$G_0W_0$ for the gaps and markedly better for the excitation energies. We also apply the method to the layered magnetic semiconductor CrI$_3$, obtaining an absorption spectrum that reproduces the main experimental features. These results establish BSE@qKI as a $GW$-free alternative to BSE@$G_0W_0$ for systematic studies of excitonic properties in complex materials.
\end{abstract}
\maketitle
%
\glsresetall

\section{Introduction}
\label{sec:intro}
Excitons --- bound states formed by electron-hole pairs --- are the quasiparticles that govern the optical response of materials~\cite{Marzari_Ferretti_Wolverton_2021}. They play a central role in several applications including photocatalysis~\cite{Balapure_rev_photocat_chargeT_2024, WANG_rev_exc_org_photocat_2026, KMICHOU2024112812}, photovoltaics~\cite{Solet_excitons_LiZnAs_ScAgC_photovoltaic_2025, Sharifzadeh_2018, Kumar_Singh_Gill_Bhattacharya_2021}, plasmonics~\cite{Yu_rev_plasmonics_2019, plasmon_exc_rev_Yi_2026}, and chemical sensing~\cite{Wolf_exc_sensing_2026}. Describing excitonic effects is therefore essential both for understanding the optical properties of materials and for designing of devices with tailored functionalities. From a first-principles perspective, \gls{mbpt}, and in particular the \gls{bse}~\cite{strinati1988application,onida_electronic_2002}, provides a state-of-the-art framework for describing neutral excitations. Conventionally, the \gls{bse} is solved using \gls{qp} energies from the $GW$ approximation~\cite{onida_electronic_2002},  often within the single-shot $G_0W_0$ scheme~\cite{Hedin_1965} (the well-known BSE@$G_0W_0$ approach). This route yields highly accurate optical spectra and excitonic properties, but its computational cost can quickly become prohibitive with material size (the scaling is typically $\geq \mathcal{O}(N^3)$, where $N$ is the number of particles) or for systematic studies over many materials~\cite{setten_automationMBPT_2017, Rasmussen_Deilmann_Thygesen_2021, bonacci2023, grosmann_robust_2024, Grunert_Grossmann_Runge_2025}. The computational burden of the $G_0W_0$ calculation is compounded by the subsequent construction and diagonalization of the \gls{bse} Hamiltonian. As such, it is desirable to decrease the cost of these calculations, without compromising the accuracy of the resulting excitonic description.

Koopmans spectral functionals~\cite{Dabo2010, Borghi2014, Nguyen2018, Linscott2023} are designed to provide accurate spectral properties at a computational cost closer to that of standard \gls{dft}~\cite{Linscott2023}, which makes them a promising source of \gls{qp} energies. By feeding their eigenvalues directly to the \gls{bse} Hamiltonian, one can bypass the $G_0W_0$ calculations altogether. We call this approach BSE@qKI. It aims to retain the predictive power of the \gls{bse} while substantially reducing the computational cost of the standard BSE@$G_0W_0$ workflow. In this work, we assess the accuracy of the BSE@qKI approach. We first benchmark it against BSE@$G_0W_0$ and against experiment for five well-characterized bulk semiconductors, and we then apply it to the layered magnetic semiconductor CrI$_3$~\cite{McGuire_Dixit_Cooper_Sales_2015}, a system in which non-collinear magnetism and spin-orbit coupling are both important. The optical spectra and excitonic properties agree well with those of BSE@$G_0W_0$; for the energy of the first excitonic peak, BSE@qKI agrees better with experiment than BSE@$G_0W_0$ does for every benchmark material considerated. The BSE@qKI approach also lends itself to automation: Koopmans functionals are implemented in the \textsc{koopmans} package~\cite{Linscott2023}, which we integrate here with the \textsc{AiiDA} workflow engine~\cite{PIZZI2016, huber2020aiida, UHRIN2021}; together, they make BSE@qKI calculations reproducible and systematic with limited user intervention. This provides a practical framework for applying the method across different classes of materials, and for carrying out high-throughput or large-scale studies of excitonic properties.

The rest of the paper is organized as follows. \Cref{sec:methods} introduces the theoretical framework underlying Koopmans functionals and the \gls{bse}, with particular emphasis on how they are combined in the BSE@qKI approach. \Cref{sec:results} illustrates the computational workflow and assesses its performance for a set of benchmark bulk semiconductors and for CrI$_3$. Finally, \cref{sec:conclusions} summarizes the main findings and discusses the prospects of Koopmans-based approaches for efficient first-principles exciton characterization. 
\section{Methods}
\label{sec:methods}
\subsection{Koopmans spectral functionals}
\label{sec:koopmans}
%
%


Koopmans spectral functionals~\cite{Dabo2010, Borghi2014, Colonna_Nguyen_Ferretti_Marzari_2018, Nguyen2018, Linscott2023} describe ground-state properties and charged excitations simultaneously and accurately through a functional approach (as opposed to a perturbative one). They give accurate spectral properties for model~\cite{schubert_testing_2023}, finite~\cite{nguyen_first-principles_2015,nguyen_first-principles_2016,elliott_koopmans_2019,Colonna_Nguyen_Ferretti_Marzari_2019, Borghi2014}, and extended systems~\cite{Colonna_Nguyen_Ferretti_Marzari_2018,de_almeida_electronic_2021,De_Genn_2022,Colonna_De_Gennaro_Linscott_Marzari_2022,ingall_accurate_2024, antimo_nicola_NC_KI, schubert_predicting_2024, stojkovic_predicting_2026} at a computational cost and complexity well below those of other state-of-the-art approaches such as quantum chemistry or Green's function methods. This is achieved by augmenting standard density-functional approximations with simple, physically motivated corrections that force the total energy to be \gls{pwl} when an electron is added to, or removed from, \textit{any} orbital in the system. 
This is an heuristic generalization --- to the entire electronic manifold --- of the well-known piecewise linearity of the total energy as a function of the number of particles~\cite{perdew_density-functional_1982} (or, equivalently, as a function of the fractional occupation of the highest occupied state). Once screening and relaxation effects are correctly accounted for~\cite{Colonna_Nguyen_Ferretti_Marzari_2018, Nguyen2018}, this restores the physical meaning of single-particle \gls{kc} energies as total-energy differences between the neutral system and the system with plus or minus one electron, and therefore amounts to a correct description of charged excitations. In practice, the generalized \gls{pwl} condition is imposed orbital by orbital: the dependence of the total energy on the occupation $f_i$ of each orbital is removed and replaced by a linear term,
\begin{equation}\label{eq:Pi_KC}
    \Pi^{\mathrm{KC}}_i = -\int_0^{f_i} \mathrm{d}s\,\langle\phi_i|\hat{H}^{\mathrm{DFT}}(s)|\phi_i\rangle + f_i\eta_i ,
\end{equation}
where $\hat{H}^{\mathrm{DFT}}(s)$ is the \gls{dft} Hamiltonian evaluated at the density in which orbital $\phi_i$ has occupation $s$. The slope $\eta_i$ of the linear term determines the flavor of the Koopmans functional~\cite{Dabo2010, Borghi2014, Linscott2023}. In the Koopmans-integral (KI) functional, $\eta_i$ is the \gls{dft} total-energy difference between the systems with orbital $i$ fully occupied and empty, $\eta_i^{\mathrm{KI}} = E^{\rm DFT}(f_i{=}1) - E^{\rm DFT}(f_i{=}0) = \int_0^1 \mathrm{d}s\,\langle\phi_i|\hat{H}^{\mathrm{DFT}}(s)|\phi_i\rangle$.
In the frozen-orbital approximation, the resulting corrections can be expressed as explicit orbital-density-dependent functionals, defining the bare (unscreened) Koopmans corrections.
Screening and relaxation effects are accounted for \emph{a posteriori} via orbital-dependent screening coefficients $\alpha_i$. The \gls{kc} total energy functional reads:
\begin{equation}\label{eq:KC_functional}
    E^{\text{KC}}[\rho, \rho_i] = E^{\text{DFT}}[\rho] + \sum_{i} \alpha_{i} \Pi^{\rm{KC}}_{i}[\rho, \rho_i],
\end{equation}
where $\rho(\mathbf{r})= \sum_i \rho_i(\mathbf{r})$ is the total charge density and $\rho_i(\mathbf{r})=f_i|\phi_i(\mathbf{r})|^2$ is the orbital density associated with orbital $\phi_i(\mathbf{r})$ at occupation $f_i$. 
In this work, we adopt the quadratic approximation to the KI functional, denoted \gls{qki}~\cite{Colonna_Nguyen_Ferretti_Marzari_2018, Colonna_De_Gennaro_Linscott_Marzari_2022}. This approximation is obtained by expanding the KI corrections to second order in $f_i$:~\cite{Colonna_Nguyen_Ferretti_Marzari_2018, Colonna_De_Gennaro_Linscott_Marzari_2022}:
\begin{align}
    \Pi_i^{\rm qKI} & = \frac{1}{2}f_i(1-f_i)\int d\mathbf{r} d\mathbf{r}' n_i(\mathbf{r}) f_{\rm Hxc} (\mathbf{r}, \mathbf{r}') n_i(\mathbf{r}'),
    \label{eq:qKI_correction}
\end{align}
where $n_i(\mathbf{r}) = |\phi_i(\mathbf{r})|^2$ is the orbital density at full occupation, and $f_{\text{Hxc}}$ is the Hartree-plus-exchange-correlation kernel --- that is, the second derivative of the underlying \gls{dft} functional with respect to the density.
The screening coefficients $\alpha_{i}$ in \cref{eq:KC_functional} account for the response of the system to the ionization process~\cite{Nguyen2018,Colonna_Nguyen_Ferretti_Marzari_2019}.
They are defined in terms of the dielectric response of the material as
 \begin{equation}
     \alpha_{i} = 
     \frac{\langle n_i | \epsilon^{-1}f_{\rm Hxc}|n_i\rangle}{\langle n_i | f_{\rm Hxc}|n_i\rangle},
\end{equation}
where $\epsilon(\mathbf{r}, \mathbf{r}')$ is the microscopic static dielectric function of the system evaluated at the underlying \gls{dft} level. 
As for other orbital-density-dependent approaches~\cite{perdew_self-interaction_1981}, \gls{kc} spectral functionals are not invariant under unitary rotations, and are minimized by a set of \textit{variational orbitals} that are typically localized in space. Imposing the generalized \gls{pwl} condition on this set of localized orbitals is what allows the method to be extended seamlessly to periodic systems~\cite{Nguyen2018}. Moreover, in most cases the variational orbitals are well approximated by \glspl{mlwf}~\cite{marzari_maximally_2012, Nguyen2018, Colonna_De_Gennaro_Linscott_Marzari_2022} that are faster to compute and which bypass the energy-minimization procedure entirely. This Koopmans-Wannier formulation~\cite{Ma_Wang_2016, Colonna_De_Gennaro_Linscott_Marzari_2022, antimo_nicola_NC_KI} is deployed as a one-shot correction to \gls{dft} eigenvalues and eigenvectors in the \verb|KCW| module of \QE~\cite{2009qe,2017qe}. In practice, one computes the matrix elements of the qKI orbital-density-dependent potential on a set of \glspl{mlwf},
\begin{equation}
    \Sigma^{\rm qKI}_{nm} = \langle \phi_n | v^{\rm qKI}_m | \phi_m\rangle,
    \label{eq:qKI_ME}
\end{equation}
where $v^{\rm qKI}_m (\mathbf{r}) = \frac{\delta \sum_j \alpha_j \Pi^{\rm qKI}_j[\rho, \rho_j]}{\delta \rho_m(\mathbf{r})}$ is the \gls{qki} potential acting on the $m$\textsuperscript{th} variational orbital. The \gls{kc} orbital-density-dependent potential $v^{\rm qKI}_m(\mathbf{r})$ acts on the variational state $|\phi_m\rangle$ as a simplified self-energy --- hence the notation in \cref{eq:qKI_ME} --- that is static and local in space~\cite{ferretti_bridging_2014}, but that includes approximate vertex corrections~\cite{Colonna_Nguyen_Ferretti_Marzari_2019, antimo_nicola_NC_KI}. 
In the standard procedure, the contribution in \cref{eq:qKI_ME} is added to the \gls{dft} Hamiltonian, represented in the same \gls{mlwf} basis, and the resulting \gls{kc} Hamiltonian is diagonalized to give the \emph{canonical orbitals}. These have a natural interpretation as Dyson orbitals, and their energies as particles additional/removal energies~\cite{ferretti_bridging_2014, nguyen_first-principles_2015, stengel_self-interaction_2008}.

%
\subsection{Bethe-Salpeter equation}
\label{sec:methods_BSE}
%

Neutral excitations arise from the creation of electron-hole pairs, which can be induced by external perturbations such as photon absorption or inelastic scattering~\cite{onida_electronic_2002}. The electron and the hole interact via the Coulomb interaction and can propagate through the system as a correlated pair; this correlated pair is the exciton.
%
Describing the optical properties of materials from an \emph{ab initio} perspective therefore requires excitonic effects to be included. Within \gls{mbpt} and the Green's functions formalism, this is achieved through the \glsentryfull{bse}~\cite{strinati1988application,onida_electronic_2002,gatti_exciton_2013}, which can be cast as an effective two-particle excitonic Hamiltonian $H^{\text{exc}}$. Within the resonant approximation and in the long-wavelength limit $\mathbf{q} \to 0$, both used throughout this work, its matrix elements read
\begin{equation}
\begin{split}
\langle vc\mathbf{k}| H^{\text{exc}} | v'c'\mathbf{k}' \rangle = 
& \left( E^{\text{qp}}_{c',\mathbf{k}'} - E^{\text{qp}}_{v',\mathbf{k}'} \right) \delta_{vc\mathbf{k},v'c'\mathbf{k}'} \\
& + \langle vc\mathbf{k} | v - W | v'c'\mathbf{k}' \rangle ,
\end{split}
\label{eq:BSE_Ham}
\end{equation}
where the electron-hole basis states are
$|vc\mathbf{k}\rangle \equiv |c,\mathbf{k}\rangle \otimes |v,\mathbf{k}\rangle$,
with $|v,\mathbf{k}\rangle$ and $|c,\mathbf{k}\rangle$ the valence and conduction single-particle states with energies $E^{\text{qp}}_{v,\mathbf{k}}$ and $E^{\text{qp}}_{c,\mathbf{k}}$. The interaction kernel consists of the bare exchange term $v$ and the screened direct interaction $W$; it is this kernel that carries the electron-hole interaction responsible for excitonic effects. The screened Coulomb interaction is calculated within the \gls{rpa}.

The \gls{bse} can be recast as an effective two-particle eigenvalue problem,
\begin{equation}
H^{\text{exc}} |\lambda\rangle = E^{\lambda} |\lambda\rangle,
\label{eq:BSE_eigenproblem}
\end{equation}
with excitonic states expanded as
\begin{equation}
|\lambda\rangle = \sum_{vc,\mathbf{k}} A^{\lambda}_{vc,\mathbf{k}} \, |vc\mathbf{k}\rangle.
\label{eq:exciton_expansion}
\end{equation}
The \gls{ebe} of a given state $|\lambda\rangle$ is usually computed as the difference between the fundamental band gap $E_g$ and the exciton eigenvalue $E^{\lambda}$. That definition is not appropriate here, however, because the lowest excitonic state may involve transitions away from the point in the \gls{bz} at which the minimum gap occurs. We therefore evaluate the \gls{ebe} as the difference between the weighted average of the independent-particle transition energies contributing to $|\lambda\rangle$, $E^{\text{qp}}_{c,\mathbf{k}} - E^{\text{qp}}_{v,\mathbf{k}}$, and the excitonic eigenvalue itself:
\begin{equation}
\text{EBE}^{\lambda} = \sum_{vc,\mathbf{k}} |A^{\lambda}_{vc,\mathbf{k}}|^2 
\left( E^{\text{qp}}_{c,\mathbf{k}} - E^{\text{qp}}_{v,\mathbf{k}} \right) - E^{\lambda}.
\label{eq:EBE}
\end{equation}
With this definition~\cite{Gao_BSEDynPPA_MoS2_2016}, the \gls{ebe} measures the red shift produced by the electron-hole interaction alone, and can be compared between different quasiparticle schemes on an equal footing.
Within this framework, the inverse macroscopic dielectric function can be expressed in terms of the excitonic eigenvalues $E^{\lambda}$ and oscillator strengths $\Omega^{\lambda}$. Considering only the resonant contributions, we obtain:
\begin{equation}
\begin{split}
\epsilon_M^{-1}(\omega) = 1 + \lim_{\textbf{q}\rightarrow 0} \frac{2}{VN_q} v(\textbf{q}) \sum_{\lambda} \Omega^{\lambda} 
\left[
\frac{1}{E^{\lambda} - (\omega + i\eta)}\right],
\end{split}
\label{eq:eps_macro}
\end{equation}
where $V$ is the volume of the unit cell and $N_q$ is the number of points in the \gls{bz} sampling. The oscillator strengths
\begin{equation}
\Omega^{\lambda} = \left| \sum_{vc,\mathbf{k}} A^{\lambda}_{vc,\mathbf{k}} 
\langle v,\mathbf{k} | e^{-i\textbf{q}\cdot \textbf{r}} | c,\mathbf{k} \rangle \right|^2
\label{eq:strength}
\end{equation}
measure the optical activity of the excitonic states and determine their contribution to the absorption spectrum.

From an experimental perspective, excitonic effects can be probed using different spectroscopic techniques depending on the transferred momentum. In the optical limit ($\mathbf{q} \to 0$) they are seen in photoabsorption measurements, while finite-momentum excitations are probed by \gls{eels} or \gls{nrixs}. These experiments give access, respectively, to the absorption spectrum Abs$(\omega)$ and the loss function $L(\omega,\mathbf{q})$, which are related to the macroscopic dielectric function by~\cite{onida_electronic_2002}
\begin{eqnarray}
\text{Abs}(\omega,\mathbf{q} \rightarrow 0) &=& \mathrm{Im}\left[\epsilon_M(\omega,\mathbf{q} \rightarrow 0) \right]
\label{eq:abs}
\\
L(\omega,\mathbf{q}) &=& -\mathrm{Im}\left[\frac{1}{\epsilon_M(\omega,\mathbf{q})} \right].
\label{eq:loss}
\end{eqnarray}
As stated above, we restrict the analysis to the optical limit; for treatments beyond the $\mathbf{q} \to 0$ regime we redirect the reader to Refs.~\citenum{gatti_exciton_2013, C3N_bonacci}.

Finally, converged results require careful control of several numerical parameters, notably the plane-wave expansion of the microscopic dielectric matrix $\epsilon_{\mathbf{G}\mathbf{G}'}(\mathbf{q},\omega)$ that enters the screened interaction $W$. We return to these parameters in \cref{sec:results-wfl}; further details are provided in Refs.~\citenum{bonacci2023, grosmann_robust_2024}.

\subsection{Combining the two: BSE@qKI}
\label{sec:methods_BSEqKI}
The quasiparticle eigenvalues $E^{\text{qp}}_{v,\mathbf{k}}$ and $E^{\text{qp}}_{c,\mathbf{k}}$ entering \cref{eq:BSE_Ham} can be obtained in several ways: from a simple scissor-and-stretching operator applied to the \gls{ks} eigenvalues, or from one-shot $G_0W_0$, i.e. the first iteration of Hedin's self-consistent loop~\cite{Hedin_1965}. The latter choice defines the standard BSE@$G_0W_0$ approach. Here we propose BSE@qKI, the direct analogue in which the $G_0W_0$ eigenvalues are replaced by \gls{qki} ones.

Making this replacement requires some care. As discussed in \cref{sec:koopmans}, diagonalizing the \gls{qki} Hamiltonian yields eigenvectors that in general differ from the \gls{dft} ones, whereas BSE@$G_0W_0$ corrects the \gls{dft} eigenvalues perturbatively and builds the \gls{bse} basis out of products of occupied and empty KS-DFT eigenvectors. As an original contribution of this work, we extend \gls{qki} to a perturbative formulation that preserves the \gls{dft} orbitals while incorporating the \gls{qki} corrections at first order. Starting from the matrix elements of the \gls{qki} potential on the variational orbitals, \cref{eq:qKI_ME}, we define the Hermitian operator
\begin{equation}
\hat{\Sigma}^{\rm qKI} = \frac{1}{2}\sum_{nm} |\phi_n \rangle \left[ \langle \phi_n | v^{\rm qKI}_m | \phi_m\rangle + \langle \phi_n | v^{\rm qKI}_n | \phi_m\rangle \right] \langle \phi_m|,
\label{eq:Sigma_qKI_operator}
\end{equation}
whose matrix elements can be computed in any basis. The perturbative \gls{qki} quasiparticle energies then follow as the first-order correction to the KS-DFT eigenvalues,
\begin{equation}
E^{\rm qKI}_i = E_i^{\rm KS} + \langle \psi^{\rm KS}_i | \hat{\Sigma}^{\rm qKI} | \psi^{\rm KS}_i \rangle, \label{eq:qKI_perturbative}
\end{equation}
where the $\psi^{\rm KS}_i$ are the KS-DFT eigenvectors of the underlying \gls{dft} energy functional. All the qKI results reported below are obtained in this perturbative form. BSE@qKI thus mirrors BSE@$G_0W_0$ step by step, while retaining the computational advantages of a Koopmans-compliant description of quasiparticle energies. Since the two schemes share the same \gls{ks} wavefunctions and the same \gls{bse} kernel, we expect them to capture the same excitonic physics and, in particular, to give comparable exciton energies; any difference in the excitation energies should then be traceable to the quasiparticle energies alone. \Cref{sec:results-validation} tests this expectation. A pioneering implementation of \gls{bse} on top of Koopmans-compliant energies was proposed by Elliott \emph{et al.}~\cite{elliott_koopmans_2019}. There, quasiparticle energies from the KI functional are used while the explicit calculation of the screening matrix $W$ is bypassed through an iterative minimization scheme on an \gls{mlwf} basis, as implemented in the \verb|Gwl| code~\cite{elliott_koopmans_2019} of \QE. That formulation was, however, limited to finite systems. Here we extend the combination of Koopmans-compliant quasiparticle energies and \gls{bse} to periodic solids and provide an automated workflow for the entire calculation, enabling its systematic application to crystalline materials. All the \gls{bse} and $G_0W_0$ calculations reported in this work were performed with the \textsc{Yambo} code~\cite{yambo1_v2,yambo2}.

\section{Results}
\label{sec:results}
\subsection{Automated workflow for BSE@qKI}
\label{sec:results-wfl}
\begin{figure*}[t]
    \centering
    \includegraphics[width=1\textwidth]{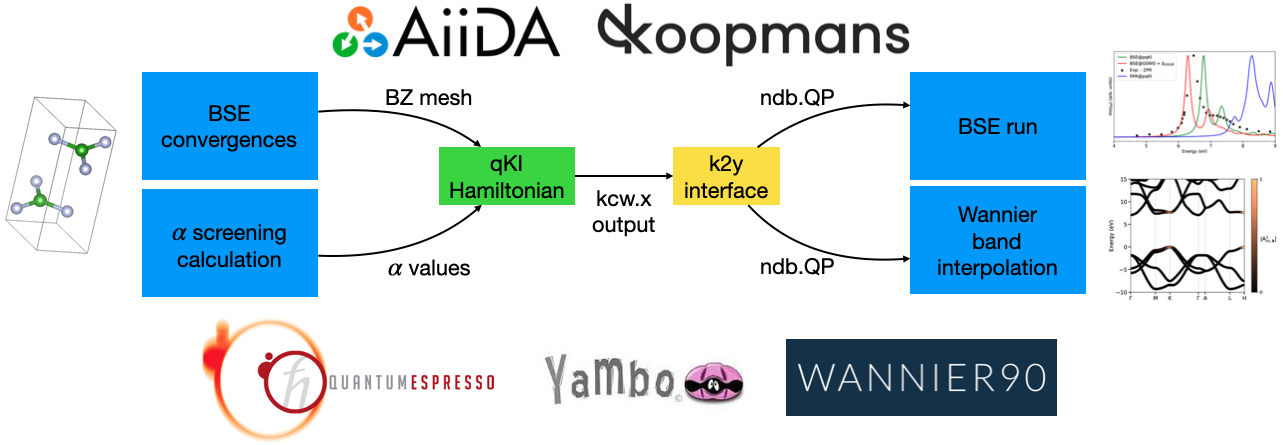}
    \caption{Schematic representation of the BSE@qKI workflow. The workflow comprises the convergence of the \glsentryshort{bse} parameters, the determination of the Koopmans screening coefficients $\{\alpha_i\}$, the calculation and conversion of qKI quasiparticle energies, and the final \glsentryshort{bse} calculation. Wannier interpolation of the qKI band structure is optionally performed as a post-processing step.}
    \label{fig:workflow}
\end{figure*}
%

The BSE@qKI workflow, sketched in \cref{fig:workflow}, consists of the following steps:
\begin{enumerate}
    \item convergence of the \gls{bse} eigenvalues with respect to relevant \textsc{Yambo} parameters;
    \item calculation of the Koopmans screening coefficients $\{\alpha_i\}$;
    \item calculation of the qKI quasiparticle eigenvalues on the converged \textbf{k}-point mesh;
    \item conversion of the Koopmans data into \textsc{Yambo}-readable format via the \texttt{k2y} interface~\cite{k2y_github};
    \item the final \gls{bse} calculation using the qKI quasiparticle energies;
    \item optionally, Wannier interpolation of the qKI band structure.
\end{enumerate}

The first step establishes the numerical parameters required for a converged \gls{bse} calculation. In particular, the excitonic energies are converged with respect to the number of bands $N_b$, the plane-wave cutoff $G_{\rm cut}$ and the $\mathbf{k}$-point sampling used for the screening matrix $W$ and the \gls{bse} kernel (\cref{eq:BSE_Ham})~\cite{bonacci2023}. The $\mathbf{k}$-point convergence is particularly important, because it controls both the sampling of the \gls{bz} and the resolution with which electron-hole transitions are represented in reciprocal space: the \gls{bse} Hamiltonian is formulated in the space of electron-hole transitions, which is defined by the selected valence and conduction bands and by the $\mathbf{k}$-point mesh. The maximum energy difference between the \gls{ks} states included in this transition space is chosen according to the spectral region of interest. Once the \gls{bse} parameters have been converged, the resulting $\mathbf{k}$-point mesh is used for the subsequent Koopmans calculation. The screening coefficients $\{\alpha_i\}$ can be determined on a coarser $\mathbf{k}$-point grid, since their convergence is independent of the numerical parameters governing the \gls{bse} calculation. The qKI Hamiltonian is then solved on the converged \gls{bse} mesh, yielding the quasiparticle eigenvalues required for the final \gls{bse} calculation. These eigenvalues are then converted into the \textsc{Yambo} format using the dedicated \texttt{k2y} interface. Starting from the original \gls{ks} wavefunctions stored in the \texttt{ns.db1} \textsc{Yambo} database and from the qKI output, \texttt{k2y} generates a new \texttt{ndb.QP} database containing the qKI quasiparticle energies. These can then be incorporated directly into the diagonal part of the \gls{bse} Hamiltonian within the standard \textsc{Yambo} workflow, without modifying the underlying \gls{ks} wavefunctions or the \gls{bse} kernel. Finally, a full \gls{bse} calculation is performed using the qKI quasiparticle energies. The \gls{bse} Hamiltonian is constructed from the qKI-corrected electron-hole transition energies and the corresponding interaction kernel; diagonalizing it provides the excitonic eigenstates and hence the optical observables, including absorption and energy-loss spectra. As an optional post-processing step, the qKI band structure can be interpolated using \glspl{mlwf}. Wannier interpolation enables the evaluation of the quasiparticle band structure along arbitrary paths in reciprocal space, including conventional high-symmetry directions. This step does not affect the \gls{bse} or the resulting optical properties, as it is performed independently on the converged qKI band structure and serves only as a means of obtaining a smooth representation of the quasiparticle dispersion.

This workflow closely parallels a conventional BSE@$G_0W_0$ calculation, but the computational cost is distributed differently. In BSE@$G_0W_0$, the cost is dominated by the $G_0W_0$ step, whose convergence depends on several coupled parameters, including the number of empty states and the plane-wave cutoff entering the calculation of the polarizability and the self-energy~\cite{bonacci2023}. In practice, the parameters required to converge the $G_0W_0$ quasiparticle energies are typically reused in the subsequent \gls{bse} calculation, which reduces the additional convergence effort associated with the \gls{bse} step; the overall cost nonetheless remains dominated by $G_0W_0$. Even when plasmon-pole approximations are employed~\cite{GN-PPA, HL-PPA}, the $GW$ calculation involves frequency-dependent quantities, whereas the \gls{bse} kernel considered here is static, so the \gls{bse} calculation is comparatively cheap. Koopmans-compliant functionals, by contrast, provide quasiparticle corrections within a variational functional framework and never require the explicit evaluation of the frequency-dependent self-energy. This makes quasiparticle energies much cheaper to obtain, while retaining --- as demonstrated in the following sections --- a level of accuracy suitable for the description of optoelectronic properties.

The complete workflow involves several electronic-structure codes: \QE\ is used for the underlying \gls{dft} and Koopmans-compliant calculations, \textsc{Wannier90}~\cite{Pizzi2020} for the construction of maximally localized Wannier functions, and \textsc{Yambo} for the many-body perturbation theory calculations. The \texttt{k2y} interface, developed within this work, provides the connection between the \textsc{koopmans} and \textsc{Yambo} workflows by converting the qKI quasiparticle energies into a \textsc{Yambo}-readable \texttt{ndb.QP} quasiparticle database. Running these steps by hand is impractical: they involve data transfer between different codes, multiple convergence studies, and the execution of a substantial number of individual calculations (typically on the order of 30 per material). Automation is therefore essential for ensuring reproducibility and for reducing the burden associated with workflow management. To this end, we employ the \textsc{AiiDA} workflow engine~\cite{PIZZI2016, huber2020aiida, UHRIN2021} and its ecosystem of plugins, including \texttt{aiida-quantumespresso}~\cite{aiida-quantumespresso}, \texttt{aiida-wannier90-workflows}~\cite{aiida_wannier90_workflows}, \texttt{aiida-yambo}~\cite{yambo2,bonacci2023}, and \texttt{aiida-yambo-wannier90}~\cite{bonacci2023}. The \texttt{yambopy} package~\cite{paleari_2025_15012963} is used for post-processing and analysis of the \textsc{Yambo} results. Koopmans-compliant calculations are managed through the \texttt{koopmans} Python package~\cite{Linscott2023}, which provides automated workflows for both the reciprocal-space implementation based on \gls{dfpt}~\cite{baroni_DFPT, Colonna_De_Gennaro_Linscott_Marzari_2022} and the finite-difference implementation based on the $\Delta$SCF approach~\cite{Nguyen2018}.
In this work, we further extend this framework through the development of the \texttt{aiida-koopmans} plugin~\cite{aiida-koopmans}, which brings Koopmans calculations into the \textsc{AiiDA} ecosystem. This provides capabilities such as remote execution, parallelization of independent tasks (\emph{e.g.}, the individual $\alpha_i$ screening calculations), and automated error handling.

The resulting workflow is reproducible and, in principle, suitable for high-throughput studies. The main user-defined inputs are the atomic structure, the Wannier projections, and the available computational resources; most of the remaining parameters can be generated automatically from predefined protocols~\cite{nascimento2025}. Two obstacles still stand in the way of complete automation. The first is the need for user-defined Wannier projections. In future work, this requirement will be removed by integrating the present workflow with recently developed automated Wannierization strategies based on projectability disentanglement and manifold remixing~\cite{qiao_projectability_2023, qiao_automated_2023}, thereby enabling the automated construction of Wannier functions spanning the target valence and conduction manifolds. The second is that, although well-established protocols exist for many \gls{dft} parameters, the convergence parameters for many-body perturbation theory calculations remain strongly system-dependent, and so must still be converged case by case. 

\subsection{Validation on well-known semiconductors}
\label{sec:results-validation}

In this section, we assess the performance of BSE@qKI for a representative set of well-known bulk semiconductors: silicon (Si), zinc oxide (ZnO), rutile titanium dioxide (TiO$_2$), diamond and hexagonal boron nitride (hBN). The central goal is to determine whether the use of qKI quasiparticle energies within the \gls{bse} framework provides excitonic properties comparable to those obtained from the established BSE@$G_0W_0$ approach. 
For each material, we first calculate the quasiparticle band structure using both qKI and $G_0W_0$, and subsequently solve the \gls{bse} on top of the corresponding quasiparticle energies. All calculations employ PseudoDojo v0.4 pseudopotentials~\cite{vanSetten2018}, within the \gls{lda} for the exchange-correlation functional. The default \QE\ parameters provided by the \texttt{aiida-quantumespresso} workflows are adopted through their predefined protocols~\cite{nascimento2025}. The parameters entering the qKI calculations are selected to ensure convergence of both the Wannier representation and the Koopmans screening coefficients, whereas the $G_0W_0$ calculations use parameters previously converged by the authors~\cite{bonacci2023}. The \gls{bse} calculations employ the same numerical parameters as the corresponding $G_0W_0$ calculations, with the exception of the Brillouin-zone sampling, whose convergence is treated separately as described in \cref{sec:results-wfl}.
Within the automated \texttt{YamboConvergence} workflow, the \gls{bse} calculations are converged by imposing a relative threshold of $\Delta\epsilon=10\%$ on the lowest excitonic eigenvalue. This threshold provides a practical compromise between computational cost and accuracy and can be tightened when higher numerical accuracy is required. In addition to the convergence of the lowest excitonic state, the optical spectrum generally requires convergence with respect to the number of valence and conduction bands included in the \gls{bse} Hamiltonian. This condition depends on the spectral range of interest and is therefore treated separately. In the present work, we focus on the low-energy region of the optical spectra and retain a limited number of low-energy transitions. Furthermore, we restrict the analysis to optically active direct excitons, also for materials whose fundamental band gap is indirect. 
The detailed computational parameters used for each material are reported in the \gls{si}. 
The purpose of this benchmark is to assess the ability of the proposed BSE@qKI approach to reproduce the results of the established BSE@$G_0W_0$ methodology across a chemically and structurally diverse set of systems. The convergence parameters adopted here provide the level of accuracy required for this comparison. For material-specific applications targeting finer, meV-scale energy differences, these parameters can be systematically tightened and a larger number of excitonic states considered.

The main results are summarized in \cref{tab:main}. For each material, we report the minimum quasiparticle band gap $E_g$ and the energy $E^1$ of the first excitonic peak. For comparison, the \gls{qsgw} band gaps are taken from the literature~\cite{chen_2015}, while the remaining theoretical results are obtained in the present work. Since no explicit BSE@QSG$\widetilde{W}$ calculations are available, the corresponding $E^1$ values are estimated by applying the QSG$\widetilde{W}$ correction $\Delta_{\text{QSG}\widetilde{W}}$ to the $E^1$ obtained from $G_0W_0$, thereby accounting for the change in the quasiparticle gap while keeping the exciton binding energy unchanged. We underline here that this is an auxiliary estimate, not a BSE calculation performed with QSG$\widetilde{W}$ energies, orbitals and screening. Its use assumes that the relevant direct transitions change approximately uniformly. Experimental band gaps are corrected for \gls{zpr} by applying the corresponding \gls{zpr} reported in the literature~\cite{Miglio_2020}. No additional \gls{zpr} correction is applied to the exciton binding energies, as previous studies have shown that their \gls{zpr} is typically substantially smaller than the electronic contribution to the band-gap renormalization~\cite{neaton_phonon_EBE_screening,Kronik_optimally_tuned_2023,Schebek_2025}.
\begin{table*}[htbp]
\centering
\small
\renewcommand{\arraystretch}{1.15}
\setlength{\tabcolsep}{6pt}

\begin{tabular}{@{}l cccc @{\hspace{2.2em}} cccc @{\hspace{2.2em}} c@{}}
\toprule
& \multicolumn{4}{c}{$E_g$ (eV)}
& \multicolumn{4}{c}{$E^1$ (eV)}
& \\
\cmidrule(lr){2-5} \cmidrule(lr){6-9}
& qKI & G$_0$W$_0$ & QSG$\tilde{W}$ & Exp.
& qKI & G$_0$W$_0$ & QSG$\tilde{W}$ & Exp.
& ZPR (eV) \\
\midrule
Silicon
& 1.32 & 1.17 & 1.30 & 1.2~\cite{chen_2015}
& 3.55 & 3.35 & 3.48 & 3.5~\cite{aspnes_dielectric_1983}
& $-0.056$ \\
ZnO
& 3.6  & 2.32 & 3.42 & 3.6~\cite{shiskin_2007}
& 3.43 & 2.16 & 3.26 & 3.5~\cite{Dario_ZnO_2024}
& $-0.16$ \\
hBN
& 7.10 & 6.15 & 6.74~\cite{cunningham_effect_2018} & 6.3~\cite{Cassabois_exp_hBN_2016}
& 6.78 & 5.51 & 6.10 & 6.3~\cite{tarrio_1989}
& $-0.22$~\cite{Shen_ZPR_hBN_2020} \\
Diamond
& 7.0  & 5.58 & 5.90 & 5.8~\cite{Miglio_2020}
& 8.13 & 6.98 & 7.30 & 8.0~\cite{Tal_tddft_2020}
& $-0.36$ \\
TiO$_2$
& 4.07 & 3.21 & 3.73 & 3.4~\cite{Miglio_2020}
& 4.73 & 3.91 & 4.43 & 4.4~\cite{Thatribud_2019}
& $-0.337$ \\
\midrule
\glsentryshort{mare} (\%)
& 13 & 10 & 6 & ---
& 4  & 16 & 4 & ---
& \\
\bottomrule
\end{tabular}

\caption{Minimum band gaps ($E_g$), first exciton energies ($E^1$), and zero-point renormalization (ZPR) for the validated semiconductors. Experimental $E_g$ and $E^1$ values are corrected for the ZPR, for comparison with theoretical results that do not include it. The $\text{QSG}\tilde{W}$ $E^1$ values are estimated by shifting the G$_0$W$_0$ spectrum to align $E_g$. The \glsentryfull{mare} is computed with experiment as reference. $\text{QSG}\tilde{W}$ and ZPR data are from Refs.~\citenum{chen_2015} and \citenum{Miglio_2020}, respectively, unless otherwise cited. Experimental $E^1$ values are extracted from plots in the cited references.}
\label{tab:main}
\end{table*}
%
We first consider the minimum quasiparticle band gaps $E_g$ reported in \cref{tab:main}. Across the five materials, qKI and $G_0W_0$ yield similar \gls{mare} with respect to experiment, respectively 13\% and 10\%, while the comparison between experiment and QSG$\widetilde{W}$ gives a \gls{mare} of 6\%. This similar average performance of qKI and $G_0W_0$, however, results from rather different material-dependent trends. In particular, qKI performs very well for ZnO, where its gap is close to both experiment and QSG$\widetilde{W}$, while $G_0W_0$ substantially underestimates it. The opposite behavior is found for diamond, where $G_0W_0$ is close to both experiment and QSG$\widetilde{W}$, whereas qKI significantly overestimates the gap.  Overall, the gap comparison does not reveal a systematic advantage of one quasiparticle approach over the other. Importantly, the MARE reported here should be regarded only as a rough comparative metric and not as a statistically significant measure of the relative performance of the different approaches. Indeed, the comparison is sensitive to individual materials: when ZnO is excluded, $G_0W_0$ shows substantially better agreement with experiment than qKI, with MAREs of 4\% and 16\%, respectively.
The observed differences reflect the distinct approximations underlying the three quasiparticle approaches. The performance of Koopmans-compliant functionals for quasiparticle energies has been investigated in a range of benchmark studies from the authors~\cite{Nguyen2018,Colonna_De_Gennaro_Linscott_Marzari_2022,De_Genn_2022}. The theoretical framework underlying the corresponding screening and quasiparticle corrections is discussed in \cref{sec:koopmans}, with further details provided in Refs.~\citenum{Colonna_Nguyen_Ferretti_Marzari_2018,Colonna_Nguyen_Ferretti_Marzari_2019,antimo_nicola_NC_KI}.

We next consider the energy $E^1$ of the first excitonic peak. For this analysis, it is important to distinguish the fundamental quasiparticle gap from the direct gap entering the \gls{bse} calculation. The former, reported in \cref{tab:main}, is the minimum gap over the Brillouin zone and can be indirect, whereas the \gls{bse} solved here describes vertical electron-hole transitions (\textbf{q}$\rightarrow$0) and therefore depends on direct quasiparticle transitions. The corresponding minimum direct gaps, $E_g^{\mathrm{dir}}$, are reported in the \gls{si}.
\begin{figure*}[htbp]
    \centering

    \begin{subfigure}{0.53\textwidth}
        \centering
        \includegraphics[width=\textwidth]{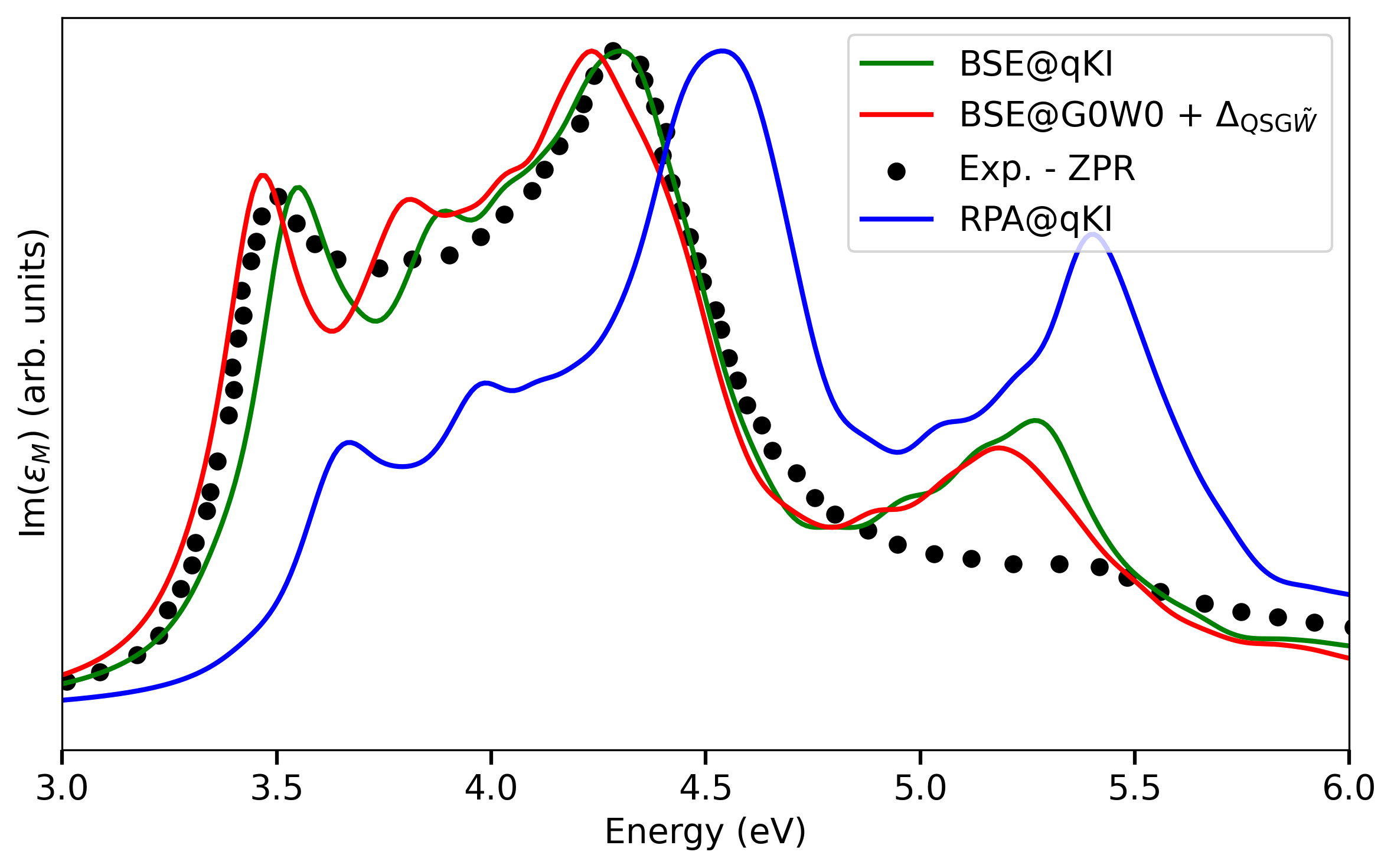}
    \end{subfigure}
    \hfill
    \begin{subfigure}{0.45\textwidth}
        \centering
        \includegraphics[width=\textwidth]{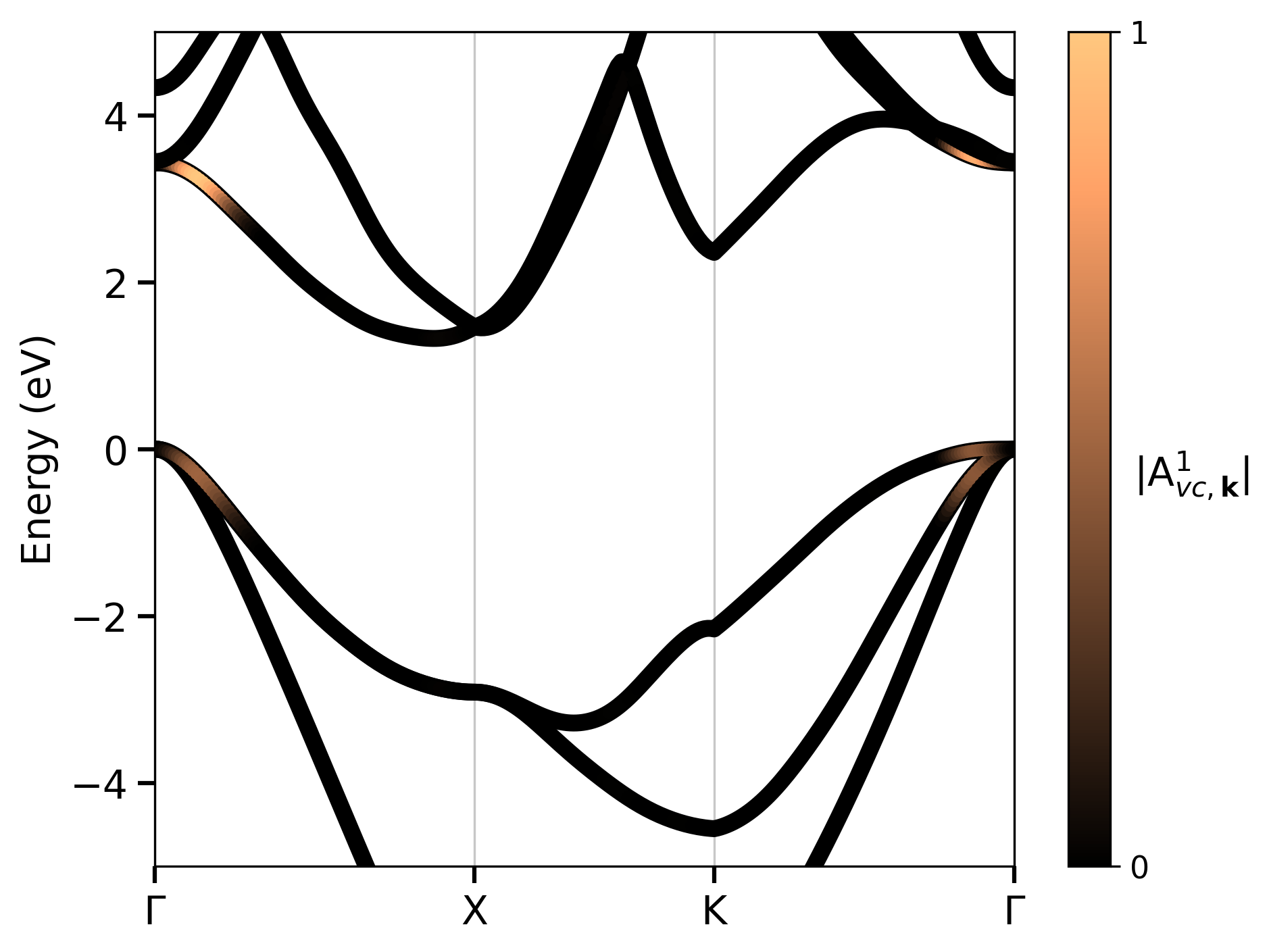}
    \end{subfigure}
    \caption{Electronic and optical properties of bulk silicon. Left: absorption spectrum obtained with BSE@qKI, BSE@$G_0W_0$, RPA@qKI and from experimental data. The BSE@$G_0W_0$ spectrum has been shifted by $\Delta_{\text{QSG}\Tilde{W}}$=130 meV to align the $G_0W_0$ gap with the QSG$\Tilde{W}$ one, while the experimental one is shifted by -\glsentryshort{zpr}=56 meV to be compared with the theoretical results. The \glsentryshort{bse} spectra are obtained with a broadening of 100 meV, comparable with the full width at half maximum of the experimental peaks. The spectra have been normalized with respect to the highest detected peak. Right: electron-hole transition amplitudes $|A_{vc,\textbf{k}}^1|$ contributing to the excitonic peak $E^1$.}
    \label{fig:Silicon_plots}
\end{figure*}
A clear improvement of BSE@qKI over BSE@$G_0W_0$ emerges for the first excitonic peak. The \gls{mare} of $E^1$ with respect to experiment is 4\% for BSE@qKI, compared with 16\% for BSE@$G_0W_0$. The corresponding QSG$\widetilde{W}$ estimate gives a \gls{mare} of 4\%. Remarkably, BSE@qKI yields a first peak closer to experiment than BSE@$G_0W_0$ for all five materials considered. The origin of this systematic difference becomes clearer when considering the direct quasiparticle gaps. For Si, qKI and $G_0W_0$ give $E_g^{\mathrm{dir}}$=3.44 and 3.25 eV, respectively, while the corresponding first excitonic peaks are 3.55 and 3.35 eV, so that the two differences, 0.19 and 0.20 eV, agree closely. For ZnO, the direct gap is 3.60 eV within qKI and 2.32 eV within $G_0W_0$, leading to first peaks at 3.43 and 2.16 eV, respectively, so that the two differences, 1.28 and 1.27 eV, again agree closely. Similarly, for hBN and diamond, the differences between the two approaches in $E^1$ closely follow those observed for the direct quasiparticle gaps: the direct gaps differ by 1.15 eV and 1.04 eV, respectively, while the corresponding differences in $E^1$ are 1.27 and 1.15 eV. A similar behavior is observed for TiO$_2$, where the direct gaps and the first peaks both differ by 0.82 eV.
These results indicate that the difference between the BSE@qKI and BSE@$G_0W_0$ spectra is predominantly determined --- as anticipated in \cref{sec:methods_BSEqKI} --- by the different quasiparticle transition energies entering the diagonal part of the \gls{bse} Hamiltonian. The electron-hole interaction kernel, by contrast, is constructed from the same underlying \gls{dft} quantities in the two approaches and therefore provides a similar excitonic renormalization (i.e. a similar \gls{ebe}, as reported in the \gls{si}). The better agreement of BSE@qKI with the optical excitation energy is connected to its direct quasiparticle transition energies rather than to a uniformly more accurate fundamental gap.

As an illustrative example, \cref{fig:Silicon_plots} shows the absorption spectrum of bulk silicon obtained with BSE@qKI and BSE@$G_0W_0$ (shifted by the corresponding $\Delta_{\text{QSG}\widetilde{W}}$), together with the experimental and RPA@qKI spectra. Both theoretical approaches reproduce the overall low-energy spectral structure, while BSE@qKI gives a first peak closer to the experimental position. The electron-hole transition amplitudes $|A_{vc,\textbf{k}}^1|$ associated with this state are shown in the right panel, providing insight into the reciprocal-space character of the corresponding excitation. The equivalent plots for the remaining materials are collected in the \gls{si}. The \glspl{ebe}, reported in the \gls{si}, are also in good agreement between qKI and $G_0W_0$, with a \gls{mare} of 10\%, consistent with the use of the same underlying \gls{dft} electronic structure in the construction of the \gls{bse} electron-hole interaction. We do not compare the calculated \glspl{ebe} with experiment at this stage, as this would require a more stringent convergence of the electron-hole transition space, particularly for excitons involving bands with high curvature near the band gap, such as in ZnO~\cite{Dario_ZnO_2024}, and for systems with an indirect fundamental gap, such as Si and hBN.
\subsection{Electronic and optical properties of bulk CrI$_3$}
\label{sec:results-cri3}
Chromium triiodide (CrI$_3$) is a prototypical two-dimensional magnetic material, and was the first  material reported to retain intrinsic ferromagnetic order down to the monolayer limit~\cite{Huang_Clark_Navarro_et_2017, Klein_MacNeill_Lado_Soriano_Navarro_2018}. Its layered structure, held together by weak van der Waals interactions,  makes it readily exfoliable and particularly attractive for spintronic and magnetoelectronic applications~\cite{Klein_MacNeill_Lado_Soriano_Navarro_2018, jiang_controlling_2018,jiang_electric-field_2018,wang_very_2018,kim_one_2018,wang_electric-field_2018,song_giant_2018,jiang_spin_2019,song_voltage_2019}. Among the chromium trihalides CrX$_3$ (X = Cl, Br, I), CrI$_3$ has the highest Curie temperature, $T_c\simeq61$~K, and a strong magnetic anisotropy~\cite{McGuire_Dixit_Cooper_Sales_2015}. Bulk CrI$_3$ also undergoes a temperature-driven structural transition, from a high-temperature monoclinic $C2/m$ phase to a low-temperature rhombohedral $R\bar{3}$ phase with ABC stacking~\cite{McGuire_Dixit_Cooper_Sales_2015}. The electronic and optical properties of CrI$_3$ have been investigated extensively using first-principles approaches~\cite{Soriano2018InterplayBI,Gibertini_2021,Acharya_Pashov_Rudenko_2022,olsen_2021,Acharya_Pashov_Cunningham_Rudenko_2021,antimo_nicola_NC_KI,Esquembre_2025}, with reported quasiparticle band gaps showing a sizeable dependence on the theoretical framework and structural configuration. In particular, QSG$\widetilde{W}$ calculations for the hexagonal hP structure have reported bulk band gaps in the range 2.2--2.5 eV~\cite{Acharya_Pashov_Rudenko_2022,Acharya_Pashov_Cunningham_Rudenko_2021}, while a Koopmans-compliant calculation for the rhombohedral $R\bar{3}$ structure reported a gap of 1.92~eV~\cite{antimo_nicola_NC_KI}. Part of this difference may therefore originate from the different structural configurations considered. 

\begin{figure*}[htbp]
    \centering

    \begin{subfigure}{0.53\textwidth}
        \centering
        \includegraphics[width=\textwidth]{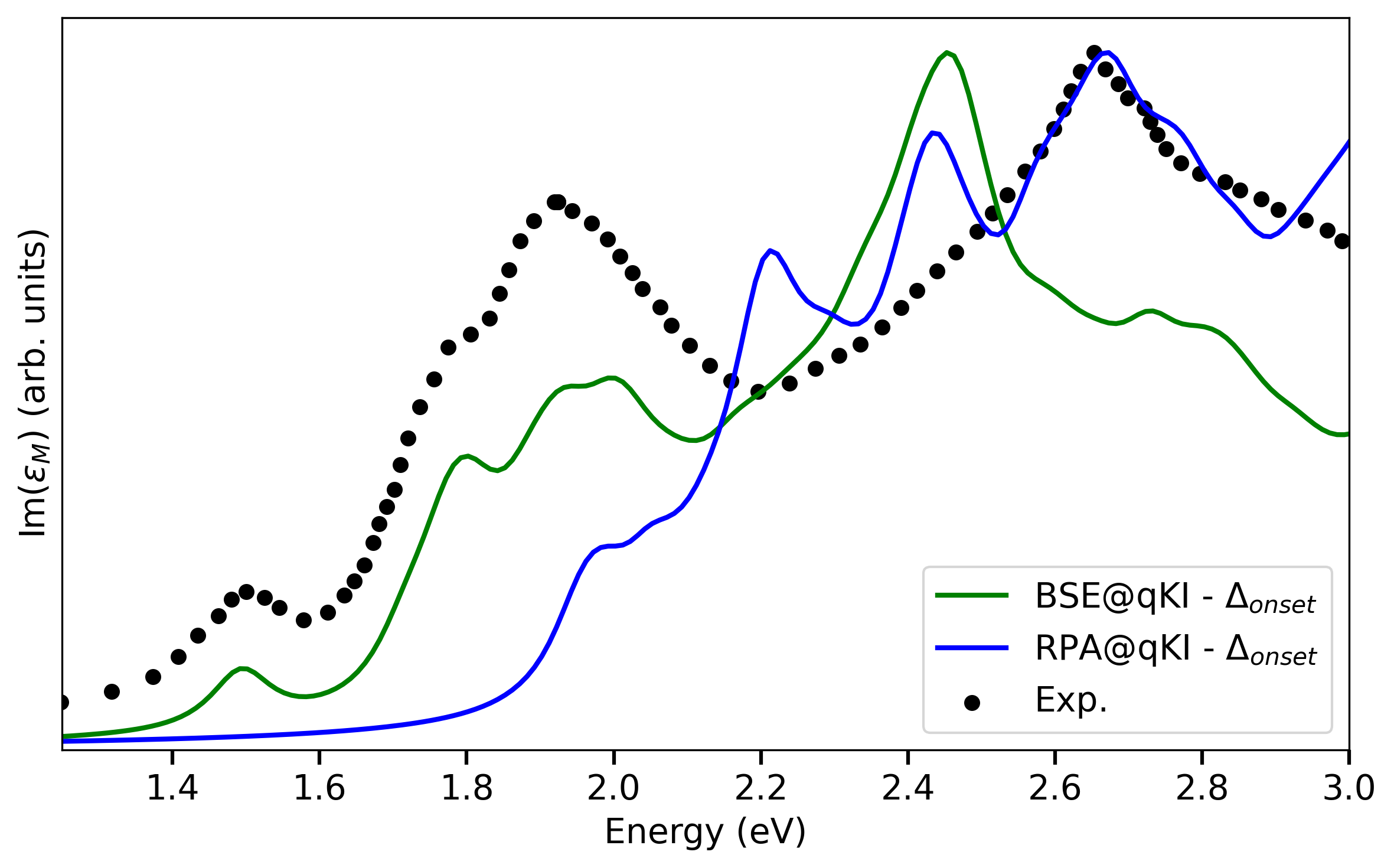}
    \end{subfigure}
    \hfill
    \begin{subfigure}{0.45\textwidth}
        \centering
        \includegraphics[width=\textwidth]{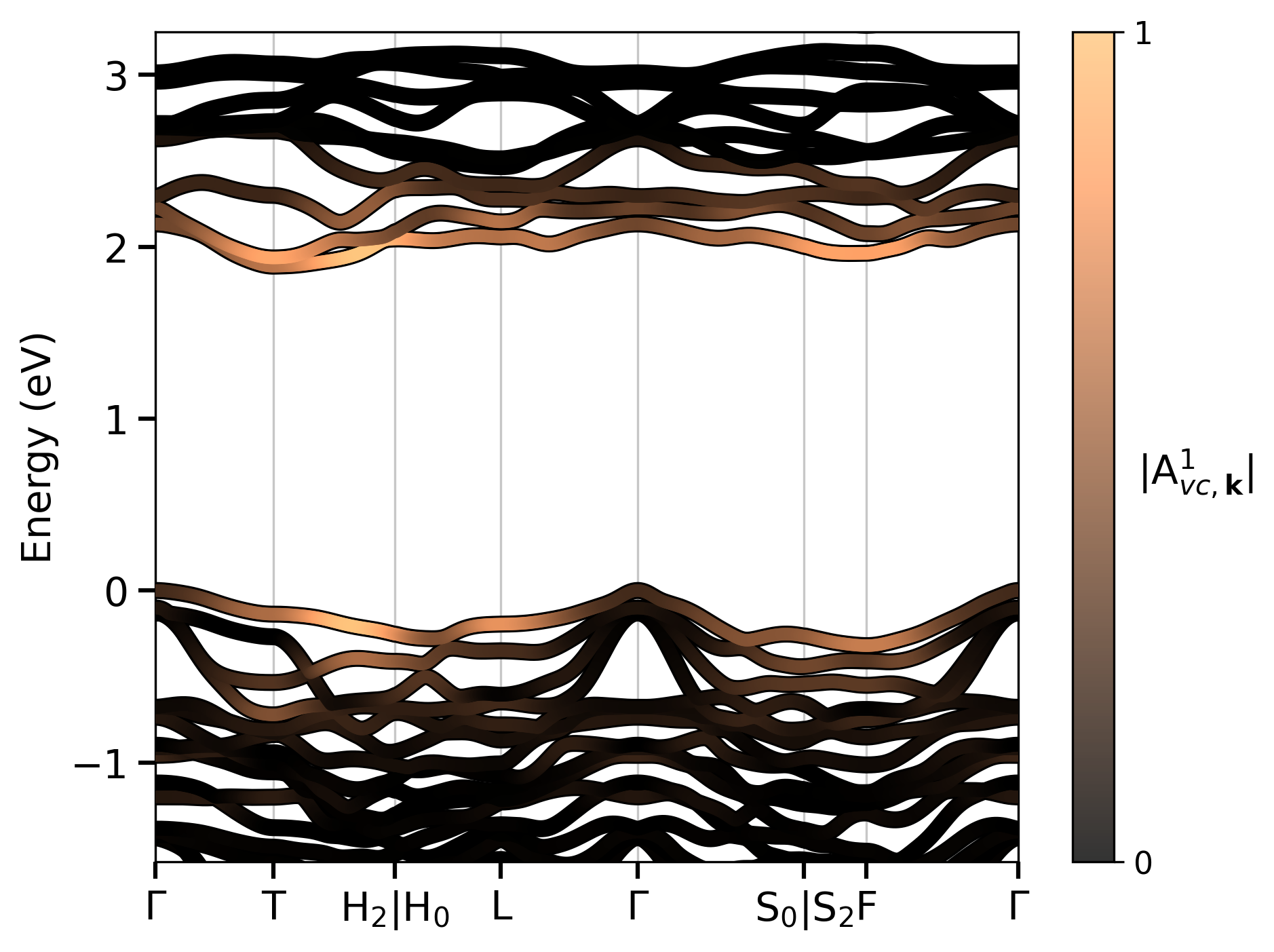}
    \end{subfigure}

    \caption{Absorption spectrum and E$^1$ exciton composition for CrI$_3$. Left: comparison between the BSE@qKI (green) and RPA (blue) absorption spectra, both rigidly shifted down by $\Delta_{\mathrm{onset}}=200$ meV, and the experimental (black) differential reflection measurements of CrI$_3$~\cite{Seyler__2018}.  The spectra have been normalized with respect to the highest included peak. Right: band structure of CrI$_3$ and electron-hole composition for the first peak of the Im($\epsilon_M$), E$^1$=1.68 eV.}
    \label{fig:double_CrI3}
\end{figure*}

In the following we focus on the $R\bar{3}$ phase: it is the low-temperature structure, and hence the one relevant to the low-temperature optical measurements against which we compare. All the simulations are performed using \gls{nc} magnetism with \gls{soc}, which has been shown to reduce the band gap by $\sim$ 300~meV~\cite{Acharya_Pashov_Rudenko_2022}.
Following the BSE@qKI approach, we first compute the band structure of CrI$_3$. At the qKI level of theory, we obtain an indirect $\Gamma \rightarrow$ T band gap of $E_g$ = 1.92 eV, as shown in the right panel of \cref{fig:double_CrI3} and in agreement with previous Koopmans predictions~\cite{antimo_nicola_NC_KI}. We then perform convergence tests for the optical properties, imposing a threshold of $\Delta \epsilon$ = 10\% on the first excitonic eigenvalue. The resulting converged set of parameters is presented in the \gls{si}.
The resulting absorption spectrum is shown in the left panel of \cref{fig:double_CrI3}, together with experimental differential reflection measurements, as presented in Ref.~\citenum{Seyler__2018}. To facilitate the comparison, the theoretical spectra are rigidly red-shifted by $\Delta_{\mathrm{onset}}=200$ meV so that its onset aligns with the experimental one. This shift should be regarded as a crude alignment rather than a quantitative correction to the theoretical spectrum; it may effectively account for effects not included in the present calculations, such as \gls{zpr} of the gap and other experimental or theoretical energy shifts. The experimental and theoretical spectra are normalized to their respective maximum intensities.
Once $\Delta_{\mathrm{onset}}$ is applied, the BSE@qKI results reproduce well the peak positions in the low-energy region of the spectrum. Structures at $\sim$ 1.5, 1.8, and 1.9–2.0 eV are correctly captured, with discrepancies smaller than 20 meV, an order of magnitude below the convergence threshold of the calculation ($\sim$160 meV). In contrast, a larger deviation of $\sim$200 meV is observed for the highest-energy peak, with the computed value at $\sim$2.43 eV compared to the experimental one at $\sim$2.65 eV. This larger discrepancy may partly reflect the more demanding convergence requirements associated with higher-energy excitonic states; a systematic description of the full high-energy spectrum, including the convergence of all relevant electron-hole transitions, is beyond the scope of the present work. Comparison with differential reflection measurements also becomes non-trivial above the continuum onset, due to the increasingly complex dependence of the reflectance on the full dielectric function, as well as possible interference and surface effects (e.g. plasmons).

The first calculated excitonic peak occurs at E$^1$=1.68~eV, which appears at 1.48~eV in the shifted spectrum of \cref{fig:double_CrI3}. The corresponding  exciton has a large binding energy, of approximately 0.5~eV. Here, as for the  indirect-gap systems discussed above, the binding energy must be referred to  the relevant direct quasiparticle transitions entering the \gls{bse} rather than to the fundamental $\Gamma\rightarrow$T gap. Its magnitude is consistent with the quasi-two-dimensional character and the weak dielectric screening associated with the layered structure of CrI$_3$. The reciprocal-space composition of this excitonic state, shown in the right panel of \cref{fig:double_CrI3}, provides further insight into its character.

\section{Conclusions}
\label{sec:conclusions}

In this work, we have developed and demonstrated an automated workflow for calculating excitonic properties within the \gls{bse} using Koopmans-compliant quasiparticle energies, thereby avoiding the computationally demanding $GW$ step typically employed in standard BSE@$G_0W_0$ calculations. 
The complete BSE@qKI workflow has been implemented within the \textsc{AiiDA} ecosystem, enabling automated execution, provenance tracking, reproducibility, and sharing of the calculations, while individual workflow components can also be used independently of \textsc{AiiDA}. As part of this work, we have developed the \texttt{k2y} interface, which provides an efficient and direct conversion between Koopmans and \textsc{Yambo} data formats and allows qKI quasiparticle energies to be incorporated into the standard \textsc{Yambo} \gls{bse} workflow.
We first validated the approach on a set of well-established bulk semiconductors. BSE@qKI provides excitonic peak energies in good agreement with experiment, and exciton binding energies in good agreement with BSE@$G_0W_0$. The differences between the two \gls{bse} schemes are primarily associated with the quasiparticle transition energies entering the diagonal part of the \gls{bse} Hamiltonian, whereas the electron-hole interaction gives a similar excitonic renormalization in both.
Finally, we applied the workflow to bulk ferromagnetic CrI$_3$, including non-collinear magnetism and spin-orbit coupling. The BSE@qKI absorption spectrum reproduces the main features of the experimental low-energy spectrum, demonstrating that the approach can be extended beyond non-magnetic benchmark semiconductors to more complex magnetic materials. The first excitonic state is characterized by a large binding energy, highlighting the relevance of electron-hole interactions in this class of layered materials.
Overall, these results establish BSE@qKI as a $GW$-free alternative to BSE@$G_0W_0$ for the calculation of excitonic and optical properties. Further work is needed to assess the performance of the approach across a broader range of materials and more demanding convergence regimes; the present workflow provides the automated and reproducible framework with which to carry out these systematic studies.
\section{Data availability}
The data presented in this work are available on the
Materials Cloud Archive~\cite{bonacci_2026_hwmbd-5t734} at the URL:
\url{https://doi.org/10.24435/materialscloud:jp-07}~\cite{bonacci_2026_hwmbd-5t734}.

\section{Code availability}
The \texttt{aiida-koopmans} plugin is openly available on GitHub at
\url{https://github.com/mikibonacci/aiida-koopmans}, and the \texttt{k2y}
code is openly available on GitHub at
\url{https://github.com/mikibonacci/k2y}.

\section{Acknowledgments}
We thank Giovanni Pizzi, Xing Wang, Jusong Yu and Nataliya Paulish for helpful discussions and valuable comments on the manuscript.
We acknowledge financial support from the NCCR MARVEL, a National Centre of Competence in Research, funded by the Swiss National Science Foundation (grant number 205602), and from the SwissTwins project, funded by the Swiss State Secretariat for Education, Research and Innovation (SERI).
We acknowledge access to Piz Daint and Alps at the Swiss National Supercomputing Centre (CSCS) under PSI's share with project ID psi15 and psi18 and under MARVEL share with project ID mr32.

\bibliographystyle{my_aip}
\bibliography{bib}

@ARTICLE{yambo1_v2,
       author = {{Marini}, Andrea and {Hogan}, Conor and {Gr{\"u}ning}, Myrta and
         {Varsano}, Daniele},
        title = "{yambo: An ab initio tool for excited state calculations}",
      journal = {Computer Physics Communications},
         year = "2009",
        month = "Aug",
       volume = {180},
        pages = {1392-1403},
          doi = {10.1016/j.cpc.2009.02.003},
archivePrefix = {arXiv},
       eprint = {0810.3118},
 primaryClass = {cond-mat.mtrl-sci},
       adsurl = {https://ui.adsabs.harvard.edu/\#abs/2009CoPhC.180.1392M}
}

@article{yambo2,
	doi = {10.1088/1361-648x/ab15d0},
	year = 2019,
	month = {may},
	publisher = {{IOP} Publishing},
	volume = {31},
	number = {32},
	pages = {325902},
	author = {D Sangalli and A Ferretti and H Miranda and C Attaccalite and I Marri and E Cannuccia and P Melo and M Marsili and F Paleari and A Marrazzo and G Prandini and P Bonf{\`{a}} and M O Atambo and F Affinito and M Palummo and A Molina-S{\'{a}}nchez and C Hogan and M Grüning and D Varsano and A Marini},
	title = {Many-body perturbation theory calculations using the yambo code},
	journal = {J. Phys.: Condens. Matter},
}

@article{bonacci2023,
  title={Towards high-throughput many-body perturbation theory: efficient algorithms and automated workflows},
  author={Bonacci, Miki and Qiao, Junfeng and Spallanzani, Nicola and Marrazzo, Antimo and Pizzi, Giovanni and Molinari, Elisa and Varsano, Daniele and Ferretti, Andrea and Prezzi, Deborah},
  journal={npj Computational Materials},
  volume={9},
  number={1},
  pages={74},
  year={2023},
  publisher={Nature Publishing Group UK London}
}

@article{Linscott2023,
    author = "Linscott, Edward B. and Colonna, Nicola and De Gennaro, Riccardo and Nguyen, Ngoc Linh and Borghi, Giovanni and Ferretti, Andrea and Dabo, Ismaila and Marzari, Nicola",
    title = "Koopmans: {{An Open-Source Package}} for {{Accurately}} and {{Efficiently Predicting Spectral Properties}} with {{Koopmans Functionals}}",
    shorttitle = "Koopmans",
    year = "2023",
    month = "August",
    journal = "J. Chem. Theory Comput.",
    publisher = "{American Chemical Society}",
    issn = "1549-9618",
    doi = "10.1021/acs.jctc.3c00652",
    urldate = "2023-08-24",
    copyright = "All rights reserved",
    note = "Introduces the koopmans code"
}

@article{De_Genn_2022,
  title = {Bloch's theorem in orbital-density-dependent functionals: Band structures from Koopmans spectral functionals},
  author = {De Gennaro, Riccardo and Colonna, Nicola and Linscott, Edward and Marzari, Nicola},
  journal = {Phys. Rev. B},
  volume = {106},
  issue = {3},
  pages = {035106},
  numpages = {14},
  year = {2022},
  month = {Jul},
  publisher = {American Physical Society},
  doi = {10.1103/PhysRevB.106.035106},
  url = {https://link.aps.org/doi/10.1103/PhysRevB.106.035106}
}

@article{Dabo2010,
    author = "Dabo, Ismaila and Ferretti, Andrea and Poilvert, Nicolas and Li, Yanli and Marzari, Nicola and Cococcioni, Matteo",
    title = "Koopmans' Condition for Density-Functional Theory",
    year = "2010",
    month = "September",
    journal = "Phys. Rev. B",
    volume = "82",
    number = "11",
    pages = "115121",
    publisher = "{American Physical Society}",
    issn = "1098-0121",
    doi = "10.1103/PhysRevB.82.115121",
    urldate = "2017-11-23",
    note = "One of the founding Koopmans functionals papers"
}

@article{Borghi2014,
    author = "Borghi, Giovanni and Ferretti, Andrea and Nguyen, Ngoc Linh and Dabo, Ismaila and Marzari, Nicola",
    title = "Koopmans-Compliant Functionals and Their Performance against Reference Molecular Data",
    year = "2014",
    month = "August",
    journal = "Phys. Rev. B",
    volume = "90",
    number = "7",
    pages = "075135",
    publisher = "{American Physical Society}",
    issn = "1098-0121",
    doi = "10.1103/PhysRevB.90.075135",
    urldate = "2017-11-23",
    note = "One of the founding Koopmans functionals papers"
}

@article{Colonna_Nguyen_Ferretti_Marzari_2018, title={Screening in Orbital-Density-Dependent Functionals}, volume={14}, ISSN={1549-9618, 1549-9626}, url={https://pubs.acs.org/doi/10.1021/acs.jctc.7b01116}, DOI={10.1021/acs.jctc.7b01116}, number={5}, journal={Journal of Chemical Theory and Computation}, author={Colonna, Nicola and Nguyen, Ngoc Linh and Ferretti, Andrea and Marzari, Nicola}, year={2018}, month=may, pages={2549–2557}, language={en} }

@article{Dario_ZnO_2024,
  title = {Unraveling many-body effects in ZnO: Combined study using momentum-resolved electron energy-loss spectroscopy and first-principles calculations},
  author = {Leon, Dario A. and Elgvin, Cana and Nguyen, Phuong Dan and Prytz, \O{}ystein and Hage, Fredrik S. and Berland, Kristian},
  journal = {Phys. Rev. B},
  volume = {109},
  issue = {11},
  pages = {115153},
  numpages = {17},
  year = {2024},
  month = {Mar},
  publisher = {American Physical Society},
  doi = {10.1103/PhysRevB.109.115153},
  url = {https://link.aps.org/doi/10.1103/PhysRevB.109.115153}
}

@article{Nguyen2018,
    author = "Nguyen, Ngoc Linh and Colonna, Nicola and Ferretti, Andrea and Marzari, Nicola",
    title = "Koopmans-Compliant Spectral Functionals for Extended Systems",
    year = "2018",
    month = "May",
    journal = "Phys. Rev. X",
    volume = "8",
    number = "2",
    pages = "021051",
    publisher = "{American Physical Society}",
    issn = "2160-3308",
    doi = "10.1103/PhysRevX.8.021051",
    urldate = "2019-03-23",
    note = "Describes Koopmans functionals in periodic systems"
}

@article{Colonna_De_Gennaro_Linscott_Marzari_2022, 
title={Koopmans Spectral Functionals in Periodic Boundary Conditions}, 
volume={18}, rights={https://doi.org/10.15223/policy-029}, ISSN={1549-9618, 1549-9626}, 
url={https://pubs.acs.org/doi/10.1021/acs.jctc.2c00161}, 
DOI={10.1021/acs.jctc.2c00161}, 
number={9}, journal={Journal of Chemical Theory and Computation}, author={Colonna, Nicola and De Gennaro, Riccardo and Linscott, Edward and Marzari, Nicola}, year={2022}, month=sept, pages={5435–5448}, language={en}
}

@article{Colonna_Nguyen_Ferretti_Marzari_2019, title={Koopmans-Compliant Functionals and Potentials and Their Application to the GW100 Test Set}, volume={15}, rights={https://doi.org/10.15223/policy-029}, ISSN={1549-9618, 1549-9626}, url={https://pubs.acs.org/doi/10.1021/acs.jctc.8b00976}, DOI={10.1021/acs.jctc.8b00976}, number={3}, journal={Journal of Chemical Theory and Computation}, author={Colonna, Nicola and Nguyen, Ngoc Linh and Ferretti, Andrea and Marzari, Nicola}, year={2019}, month=mar, pages={1905–1914}, language={en} }

@article{antimo_nicola_NC_KI,
  title = {Spin-dependent interactions in orbital-density-dependent functionals: Noncollinear Koopmans spectral functionals},
  author = {Marrazzo, Antimo and Colonna, Nicola},
  journal = {Phys. Rev. Res.},
  volume = {6},
  issue = {3},
  pages = {033085},
  numpages = {18},
  year = {2024},
  month = {Jul},
  publisher = {American Physical Society},
  doi = {10.1103/PhysRevResearch.6.033085},
  url = {https://link.aps.org/doi/10.1103/PhysRevResearch.6.033085}
}

@article{vanSetten2018,
    author = "{van Setten}, M. J. and Giantomassi, M. and Bousquet, E. and Verstraete, M. J. and Hamann, D. R. and Gonze, X. and Rignanese, G. -M.",
    title = "The {{PseudoDojo}}: {{Training}} and Grading a 85 Element Optimized Norm-Conserving Pseudopotential Table",
    shorttitle = "The {{PseudoDojo}}",
    year = "2018",
    month = "May",
    journal = "Computer Physics Communications",
    volume = "226",
    pages = "39--54",
    issn = "0010-4655",
    doi = "10.1016/j.cpc.2018.01.012",
    urldate = "2022-07-29",
    langid = "english",
    note = "Citation for the pseudo dojo standard v0.4.1 pseudopotential library"
}

@ARTICLE{2009qe,
       author = {{Giannozzi}, Paolo and {Baroni}, Stefano and {Bonini}, Nicola and {Caland
        ra}, Matteo and {Car}, Roberto and {Cavazzoni}, Carlo and
         {Ceresoli}, Davide and {Chiarotti}, Guido L. and {Cococcioni}, Matteo and
         {Dabo}, Ismaila and {Dal Corso}, Andrea and {de Gironcoli}, Stefano and
         {Fabris}, Stefano and {Fratesi}, Guido and {Gebauer}, Ralph and
         {Gerstmann}, Uwe and {Gougoussis}, Christos and {Kokalj}, Anton and
         {Lazzeri}, Michele and {Martin-Samos}, Layla and {Marzari}, Nicola and
         {Mauri}, Francesco and {Mazzarello}, Riccardo and {Paolini}, Stefano and
         {Pasquarello}, Alfredo and {Paulatto}, Lorenzo and {Sbraccia}, Carlo and
         {Scandolo}, Sandro and {Sclauzero}, Gabriele and {Seitsonen}, Ari P. and
         {Smogunov}, Alexander and {Umari}, Paolo and {Wentzcovitch}, Renata M.},
        title = "{QUANTUM ESPRESSO: a modular and open-source software project for quantum simulations of materials}",
      journal = {Journal of Physics Condensed Matter},
         year = "2009",
        month = "Sep",
       volume = {21},
          eid = {395502},
        pages = {395502},
          doi = {10.1088/0953-8984/21/39/395502},
archivePrefix = {arXiv},
       eprint = {0906.2569},
 primaryClass = {cond-mat.mtrl-sci},
       adsurl = {https://ui.adsabs.harvard.edu/\#abs/2009JPCM...21M5502G}
}

@ARTICLE{2017qe,
       author = {{Giannozzi}, P. and {Andreussi}, O. and {Brumme}, T. and {Bunau}, O. and
         {Buongiorno Nardelli}, M. and {Calandra}, M. and {Car}, R. and
         {Cavazzoni}, C. and {Ceresoli}, D. and {Cococcioni}, M. and
         {Colonna}, N. and {Carnimeo}, I. and {Dal Corso}, A. and
         {de Gironcoli}, S. and {Delugas}, P. and {DiStasio}, R.~A., Jr. and
         {Ferretti}, A. and {Floris}, A. and {Fratesi}, G. and {Fugallo}, G. and
         {Gebauer}, R. and {Gerstmann}, U. and {Giustino}, F. and {Gorni}, T. and
         {Jia}, J. and {Kawamura}, M. and {Ko}, H. -Y. and {Kokalj}, A. and
         {K{\"u}{\c{c}}{\"u}kbenli}, E. and {Lazzeri}, M. and {Marsili}, M. and
         {Marzari}, N. and {Mauri}, F. and {Nguyen}, N.~L. and {Nguyen}, H. -V. and
         {Otero-de-la-Roza}, A. and {Paulatto}, L. and {Ponc{\'e}}, S. and
         {Rocca}, D. and {Sabatini}, R. and {Santra}, B. and {Schlipf}, M. and
         {Seitsonen}, A.~P. and {Smogunov}, A. and {Timrov}, I. and
         {Thonhauser}, T. and {Umari}, P. and {Vast}, N. and {Wu}, X. and
         {Baroni}, S.},
        title = "{Advanced capabilities for materials modelling with Quantum ESPRESSO}",
      journal = {Journal of Physics Condensed Matter},
         year = "2017",
        month = "Nov",
       volume = {29},
          eid = {465901},
        pages = {465901},
          doi = {10.1088/1361-648X/aa8f79},
archivePrefix = {arXiv},
       eprint = {1709.10010},
 primaryClass = {cond-mat.mtrl-sci},
       adsurl = {https://ui.adsabs.harvard.edu/\#abs/2017JPCM...29T5901G}
}

@article{Pizzi2020,
	doi = {10.1088/1361-648x/ab51ff},
	url = {https://doi.org/10.1088%2F1361-648x%2Fab51ff},
	year = 2020,
	month = {jan},
	publisher = {{IOP} Publishing},
	volume = {32},
	number = {16},
	pages = {165902},
	author = {Giovanni Pizzi and Valerio Vitale and Ryotaro Arita and Stefan Blügel and Frank Freimuth and Guillaume G{\'{e}}ranton and Marco Gibertini and Dominik Gresch and Charles Johnson and Takashi Koretsune and Julen Iba{\~{n}}ez-Azpiroz and Hyungjun Lee and Jae-Mo Lihm and Daniel Marchand and Antimo Marrazzo and Yuriy Mokrousov and Jamal I Mustafa and Yoshiro Nohara and Yusuke Nomura and Lorenzo Paulatto and Samuel Ponc{\'{e}} and Thomas Ponweiser and Junfeng Qiao and Florian Thöle and Stepan S Tsirkin and Ma{\l}gorzata Wierzbowska and Nicola Marzari and David Vanderbilt and Ivo Souza and Arash A Mostofi and Jonathan R Yates},
	title = {Wannier90 as a community code: new features and applications},
	journal = {Journal of Physics Condensed Matter}
}

@software{paleari_2025_15012963,
  author       = {Paleari, Fulvio and
                  Molina-Sánchez, Alejandro and
                  Nalabothula, Muralidhar and
                  Reho, Riccardo and
                  Bonacci, Miki and
                  Castelo, José and
                  Cervantes-Villanueva, Jorge and
                  Pionteck, Mike and
                  Silvetti, Martino and
                  Attaccalite, Claudio and
                  Pereira Coutada Miranda, Henrique},
  title        = {Yambopy},
  month        = mar,
  year         = 2025,
  publisher    = {Zenodo},
  version      = {0.4.0},
  doi          = {10.5281/zenodo.15012963},
  url          = {https://doi.org/10.5281/zenodo.15012963},
  swhid        = {swh:1:dir:f32478c6a093399c0265433dca241b041fc8d849
                   ;origin=https://doi.org/10.5281/zenodo.15012962;vi
                   sit=swh:1:snp:d8aeaebeb724d88b71fbd3ff67863307aa9e
                   145b;anchor=swh:1:rel:376827034856872ee43e025008b4
                   4102c5f5b87b;path=yambopy-0.4.0
                  },
}

@article{cunningham_effect_2018,
	title = {Effect of ladder diagrams on optical absorption spectra in a quasiparticle self-consistent {GW} framework},
	volume = {2},
	issn = {2475-9953},
	url = {https://link.aps.org/doi/10.1103/PhysRevMaterials.2.034603},
	doi = {10.1103/PhysRevMaterials.2.034603},
	language = {en},
	number = {3},
	urldate = {2025-11-19},
	journal = {Physical Review Materials},
	author = {Cunningham, Brian and Grüning, Myrta and Azarhoosh, Pooya and Pashov, Dimitar and Van Schilfgaarde, Mark},
	month = mar,
	year = {2018},
	pages = {034603},
}

@article{aspnes_dielectric_1983,
	title = {Dielectric functions and optical parameters of {Si}, {Ge}, {GaP}, {GaAs}, {GaSb}, {InP}, {InAs}, and {InSb} from 1.5 to 6.0 {eV}},
	volume = {27},
	copyright = {http://link.aps.org/licenses/aps-default-license},
	issn = {0163-1829},
	url = {https://link.aps.org/doi/10.1103/PhysRevB.27.985},
	doi = {10.1103/PhysRevB.27.985},
	language = {en},
	number = {2},
	urldate = {2025-11-19},
	journal = {Physical Review B},
	author = {Aspnes, D. E. and Studna, A. A.},
	month = jan,
	year = {1983},
	pages = {985--1009},
}

@article{nascimento2025,
      title={Accurate and efficient protocols for high-throughput first-principles materials simulations}, 
      author={Gabriel de Miranda Nascimento and Flaviano Jos\'e dos Santos and Marnik Bercx and Davide Grassano and Giovanni Pizzi and Nicola Marzari},
      year={2025},
      volume={2504.03962},
      journal={arXiv}
  }

@article{Louie_TiO2,
  title = {First-principles $\mathrm{DFT}+GW$ study of oxygen vacancies in rutile ${\text{TiO}}_{2}$},
  author = {Malashevich, Andrei and Jain, Manish and Louie, Steven G.},
  journal = {Phys. Rev. B},
  volume = {89},
  issue = {7},
  pages = {075205},
  numpages = {7},
  year = {2014},
  month = {Feb},
  publisher = {American Physical Society},
  doi = {10.1103/PhysRevB.89.075205},
  url = {https://link.aps.org/doi/10.1103/PhysRevB.89.075205}
}

@article{McGuire_Dixit_Cooper_Sales_2015, title={Coupling of Crystal Structure and Magnetism in the Layered, Ferromagnetic Insulator CrI3}, volume={27}, ISSN={0897-4756, 1520-5002}, url={https://pubs.acs.org/doi/10.1021/cm504242t}, DOI={10.1021/cm504242t}, number={2}, journal={Chemistry of Materials}, author={McGuire, Michael A. and Dixit, Hemant and Cooper, Valentino R. and Sales, Brian C.}, year={2015}, month=jan, pages={612–620}, language={en} }

@article{Huang_Clark_Navarro_et_2017, title={Layer-dependent ferromagnetism in a van der Waals crystal down to the monolayer limit}, volume={546}, ISSN={0028-0836, 1476-4687}, url={https://www.nature.com/articles/nature22391}, DOI={10.1038/nature22391}, number={7657}, journal={Nature}, author={Huang, Bevin and Clark, Genevieve and Navarro-Moratalla, Efrén and Klein, Dahlia R. and Cheng, Ran and Seyler, Kyle L. and Zhong, Ding and Schmidgall, Emma and McGuire, Michael A. and Cobden, David H. and Yao, Wang and Xiao, Di and Jarillo-Herrero, Pablo and Xu, Xiaodong}, year={2017}, month=june, pages={270–273}, language={en} }

@article{Klein_MacNeill_Lado_Soriano_Navarro_2018, title={Probing magnetism in 2D van der Waals crystalline insulators via electron tunneling}, volume={360}, ISSN={0036-8075, 1095-9203}, url={https://www.science.org/doi/10.1126/science.aar3617}, DOI={10.1126/science.aar3617}, number={6394}, journal={Science}, author={Klein, D. R. and MacNeill, D. and Lado, J. L. and Soriano, D. and Navarro-Moratalla, E. and Watanabe, K. and Taniguchi, T. and Manni, S. and Canfield, P. and Fernández-Rossier, J. and Jarillo-Herrero, P.}, year={2018}, month=june, pages={1218–1222}, language={en} }

@article{jiang_controlling_2018,
	title = {Controlling magnetism in {2D} {CrI3} by electrostatic doping},
	volume = {13},
	issn = {1748-3387, 1748-3395},
	url = {https://www.nature.com/articles/s41565-018-0135-x},
	doi = {10.1038/s41565-018-0135-x},
	language = {en},
	number = {7},
	urldate = {2026-03-20},
	journal = {Nature Nanotechnology},
	author = {Jiang, Shengwei and Li, Lizhong and Wang, Zefang and Mak, Kin Fai and Shan, Jie},
	month = jul,
	year = {2018},
	pages = {549--553},
}

@article{jiang_electric-field_2018,
	title = {Electric-field switching of two-dimensional van der {Waals} magnets},
	volume = {17},
	issn = {1476-1122, 1476-4660},
	url = {https://www.nature.com/articles/s41563-018-0040-6},
	doi = {10.1038/s41563-018-0040-6},
	language = {en},
	number = {5},
	urldate = {2026-03-20},
	journal = {Nature Materials},
	author = {Jiang, Shengwei and Shan, Jie and Mak, Kin Fai},
	month = may,
	year = {2018},
	pages = {406--410},
}

@article{wang_electric-field_2018,
	title = {Electric-field control of magnetism in a few-layered van der {Waals} ferromagnetic semiconductor},
	volume = {13},
	issn = {1748-3387, 1748-3395},
	url = {https://www.nature.com/articles/s41565-018-0186-z},
	doi = {10.1038/s41565-018-0186-z},
	language = {en},
	number = {7},
	urldate = {2026-03-20},
	journal = {Nature Nanotechnology},
	author = {Wang, Zhi and Zhang, Tongyao and Ding, Mei and Dong, Baojuan and Li, Yanxu and Chen, Maolin and Li, Xiaoxi and Huang, Jianqi and Wang, Hanwen and Zhao, Xiaotian and Li, Yong and Li, Da and Jia, Chuankun and Sun, Lidong and Guo, Huaihong and Ye, Yu and Sun, Dongming and Chen, Yuansen and Yang, Teng and Zhang, Jing and Ono, Shimpei and Han, Zheng and Zhang, Zhidong},
	month = jul,
	year = {2018},
	pages = {554--559},
}

@article{jiang_spin_2019,
	title = {Spin tunnel field-effect transistors based on two-dimensional van der {Waals} heterostructures},
	volume = {2},
	issn = {2520-1131},
	url = {https://www.nature.com/articles/s41928-019-0232-3},
	doi = {10.1038/s41928-019-0232-3},
	language = {en},
	number = {4},
	urldate = {2026-03-20},
	journal = {Nature Electronics},
	author = {Jiang, Shengwei and Li, Lizhong and Wang, Zefang and Shan, Jie and Mak, Kin Fai},
	month = apr,
	year = {2019},
	pages = {159--163},
}

@article{song_giant_2018,
	title = {Giant tunneling magnetoresistance in spin-filter van der {Waals} heterostructures},
	volume = {360},
	issn = {0036-8075, 1095-9203},
	url = {https://www.science.org/doi/10.1126/science.aar4851},
	doi = {10.1126/science.aar4851},
	language = {en},
	number = {6394},
	urldate = {2026-03-20},
	journal = {Science},
	author = {Song, Tiancheng and Cai, Xinghan and Tu, Matisse Wei-Yuan and Zhang, Xiaoou and Huang, Bevin and Wilson, Nathan P. and Seyler, Kyle L. and Zhu, Lin and Taniguchi, Takashi and Watanabe, Kenji and McGuire, Michael A. and Cobden, David H. and Xiao, Di and Yao, Wang and Xu, Xiaodong},
	month = jun,
	year = {2018},
	pages = {1214--1218},
}

@article{song_voltage_2019,
	title = {Voltage {Control} of a van der {Waals} {Spin}-{Filter} {Magnetic} {Tunnel} {Junction}},
	volume = {19},
	copyright = {https://doi.org/10.15223/policy-029},
	issn = {1530-6984, 1530-6992},
	url = {https://pubs.acs.org/doi/10.1021/acs.nanolett.8b04160},
	doi = {10.1021/acs.nanolett.8b04160},
	language = {en},
	number = {2},
	urldate = {2026-03-20},
	journal = {Nano Letters},
	author = {Song, Tiancheng and Tu, Matisse Wei-Yuan and Carnahan, Caitlin and Cai, Xinghan and Taniguchi, Takashi and Watanabe, Kenji and McGuire, Michael A. and Cobden, David H. and Xiao, Di and Yao, Wang and Xu, Xiaodong},
	month = feb,
	year = {2019},
	pages = {915--920},
}

@article{wang_very_2018,
	title = {Very large tunneling magnetoresistance in layered magnetic semiconductor {CrI3}},
	volume = {9},
	issn = {2041-1723},
	url = {https://www.nature.com/articles/s41467-018-04953-8},
	doi = {10.1038/s41467-018-04953-8},
	number = {1},
	urldate = {2026-03-20},
	journal = {Nature Communications},
	author = {Wang, Zhe and Gutiérrez-Lezama, Ignacio and Ubrig, Nicolas and Kroner, Martin and Gibertini, Marco and Taniguchi, Takashi and Watanabe, Kenji and Imamoğlu, Ataç and Giannini, Enrico and Morpurgo, Alberto F.},
	month = jun,
	year = {2018},
	pages = {2516},
}

@article{kim_one_2018,
	title = {One {Million} {Percent} {Tunnel} {Magnetoresistance} in a {Magnetic} van der {Waals} {Heterostructure}},
	volume = {18},
	issn = {1530-6984, 1530-6992},
	url = {https://pubs.acs.org/doi/10.1021/acs.nanolett.8b01552},
	doi = {10.1021/acs.nanolett.8b01552},
	language = {en},
	number = {8},
	urldate = {2026-03-20},
	journal = {Nano Letters},
	author = {Kim, Hyun Ho and Yang, Bowen and Patel, Tarun and Sfigakis, Francois and Li, Chenghe and Tian, Shangjie and Lei, Hechang and Tsen, Adam W.},
	month = aug,
	year = {2018},
	pages = {4885--4890},
}

@article{Acharya_Pashov_Cunningham_Rudenko_2021, title={Electronic structure of chromium trihalides beyond density functional theory}, volume={104}, ISSN={2469-9950, 2469-9969}, url={https://link.aps.org/doi/10.1103/PhysRevB.104.155109}, DOI={10.1103/PhysRevB.104.155109}, number={15}, journal={Physical Review B}, author={Acharya, Swagata and Pashov, Dimitar and Cunningham, Brian and Rudenko, Alexander N. and Rösner, Malte and Grüning, Myrta and Van Schilfgaarde, Mark and Katsnelson, Mikhail I.}, year={2021}, month=oct, pages={155109}, language={en} }

@article{Acharya_Pashov_Rudenko_2022, title={Real- and momentum-space description of the excitons in bulk and monolayer chromium tri-halides}, volume={6}, ISSN={2397-7132}, url={https://www.nature.com/articles/s41699-022-00307-7}, DOI={10.1038/s41699-022-00307-7}, number={1}, journal={npj 2D Materials and Applications}, author={Acharya, Swagata and Pashov, Dimitar and Rudenko, Alexander N. and Rösner, Malte and Schilfgaarde, Mark Van and Katsnelson, Mikhail I.}, year={2022}, month=may, pages={33}, language={en} }

@article{Seyler__2018, title={Ligand-field helical luminescence in a 2D ferromagnetic insulator}, volume={14}, ISSN={1745-2473, 1745-2481}, url={https://www.nature.com/articles/s41567-017-0006-7}, DOI={10.1038/s41567-017-0006-7}, number={3}, journal={Nature Physics}, author={Seyler, Kyle L. and Zhong, Ding and Klein, Dahlia R. and Gao, Shiyuan and Zhang, Xiaoou and Huang, Bevin and Navarro-Moratalla, Efrén and Yang, Li and Cobden, David H. and McGuire, Michael A. and Yao, Wang and Xiao, Di and Jarillo-Herrero, Pablo and Xu, Xiaodong}, year={2018}, month=mar, pages={277–281}, language={en} }

@article{schubert_testing_2023,
	title = {Testing {Koopmans} spectral functionals on the analytically solvable {Hooke}’s atom},
	volume = {158},
	issn = {0021-9606},
	url = {https://doi.org/10.1063/5.0138610},
	doi = {10.1063/5.0138610},
	number = {14},
	urldate = {2025-11-26},
	journal = {The Journal of Chemical Physics},
	author = {Schubert, Yannick and Marzari, Nicola and Linscott, Edward},
	month = apr,
	year = {2023},
	pages = {144113},
}

@article{nguyen_first-principles_2015,
	title = {{First-Principles Photoemission Spectroscopy and Orbital Tomography in Molecules from Koopmans-Compliant Functionals}},
	volume = {114},
	doi = {10.1103/PhysRevLett.114.166405},
	pages = {166405},
	number = {16},
	journal = {Physical Review Letters},
	author = {Nguyen, Ngoc Linh and Borghi, Giovanni and Ferretti, Andrea and Dabo, Ismaila and Marzari, Nicola},
    year = {2015},
}

@article{nguyen_first-principles_2016,
	title = {{First-Principles Photoemission Spectroscopy of} {DNA} and {RNA} {Nucleobases} from {Koopmans-Compliant Functionals}},
	volume = {12},
	issn = {1549-9618},
	doi = {10.1021/acs.jctc.6b00145},
	pages = {3948--3958},
	number = {8},
	journal = {Journal of Chemical Theory and Computation},
	author = {Nguyen, Ngoc Linh and Borghi, Giovanni and Ferretti, Andrea and Marzari, Nicola},
  year = 2016
}

@article{elliott_koopmans_2019,
author = {Elliott, Joshua D. and Colonna, Nicola and Marsili, Margherita and Marzari, Nicola and Umari, Paolo},
title = {{Koopmans Meets Bethe--Salpeter: Excitonic Optical Spectra without GW}},
journal = {Journal of Chemical Theory and Computation},
volume = {15},
number = {6},
pages = {3710-3720},
year = {2019},
doi = {10.1021/acs.jctc.8b01271},
}

@article{de_almeida_electronic_2021,
	title = {{Electronic Structure of Water from Koopmans-Compliant Functionals}},
	volume = {17},
	issn = {1549-9618},
	doi = {10.1021/acs.jctc.1c00063},
	pages = {3923--3930},
	number = {7},
    year = {2021},
	journal = {Journal of Chemical Theory and Computation},
	author = {de Almeida, James Moraes and Nguyen, Ngoc Linh and Colonna, Nicola and Chen, Wei and Rodrigues Miranda, Caetano and Pasquarello, Alfredo and Marzari, Nicola},
}

@article{ingall_accurate_2024,
	title = {Accurate and {Efficient} {Computation} of the {Fundamental} {Bandgap} of the {Vacancy}-{Ordered} {Double} {Perovskite} {Cs2TiBr6}},
	volume = {128},
	issn = {1932-7447},
	url = {https://doi.org/10.1021/acs.jpcc.3c07957},
	doi = {10.1021/acs.jpcc.3c07957},
	number = {22},
	urldate = {2024-06-28},
	journal = {The Journal of Physical Chemistry C},
	author = {Ingall, J. E. and Linscott, Edward and Colonna, Nicola and Page, Alister J. and Keast, Vicki J.},
	month = jun,
	year = {2024},
	note = {Publisher: American Chemical Society},
	pages = {9217--9228},
}

@article{schubert_predicting_2024,
	title = {Predicting electronic screening for fast {Koopmans} spectral functional calculations},
	volume = {10},
	copyright = {2024 The Author(s)},
	issn = {2057-3960},
	url = {https://www.nature.com/articles/s41524-024-01484-3},
	doi = {10.1038/s41524-024-01484-3},
	language = {en},
	number = {1},
	urldate = {2025-11-26},
	journal = {npj Computational Materials},
	author = {Schubert, Yannick and Luber, Sandra and Marzari, Nicola and Linscott, Edward},
	month = dec,
	year = {2024},
	note = {Publisher: Nature Publishing Group},
	pages = {299},
}

@article{stojkovic_predicting_2026,
	title = {Predicting the suitability of photocatalysts for water splitting using {Koopmans} spectral functionals: {The} case of TiO$_2$ polymorphs},
	volume = {10},
	shorttitle = {Predicting the suitability of photocatalysts for water splitting using {Koopmans} spectral functionals},
	url = {https://link.aps.org/doi/10.1103/6gqw-thc1},
	doi = {10.1103/6gqw-thc1},
	number = {2},
	urldate = {2026-06-15},
	journal = {Physical Review Materials},
	publisher = {American Physical Society},
	author = {Stojkovic, Marija and Linscott, Edward and Marzari, Nicola},
	month = feb,
	year = {2026},
	pages = {025403},
}

@article{perdew_density-functional_1982,
	title = {Density-{Functional} {Theory} for {Fractional} {Particle} {Number}: {Derivative} {Discontinuities} of the {Energy}},
	volume = {49},
	shorttitle = {Density-{Functional} {Theory} for {Fractional} {Particle} {Number}},
	url = {http://link.aps.org/doi/10.1103/PhysRevLett.49.1691},
	doi = {10.1103/PhysRevLett.49.1691},
	number = {23},
	urldate = {2015-10-04},
	journal = {Physical Review Letters},
	author = {Perdew, John P. and Parr, Robert G. and Levy, Mel and Balduz, Jose L.},
	year = {1982},
	pages = {1691--1694},
}

@article{marzari_maximally_2012,
	title = {Maximally localized {Wannier} functions: {Theory} and applications},
	volume = {84},
	shorttitle = {Maximally localized {Wannier} functions},
	url = {http://link.aps.org/doi/10.1103/RevModPhys.84.1419},
	doi = {10.1103/RevModPhys.84.1419},
	urldate = {2017-03-28},
	journal = {Rev. Mod. Phys.},
	author = {Marzari, Nicola and Mostofi, Arash A. and Yates, Jonathan R. and Souza, Ivo and Vanderbilt, David},
	month = oct,
	year = {2012},
	pages = {1419--1475}
}

@article{ferretti_bridging_2014,
	title = {Bridging density-functional and many-body perturbation theory: {Orbital}-density dependence in electronic-structure functionals},
	volume = {89},
	shorttitle = {Bridging density-functional and many-body perturbation theory},
	url = {http://link.aps.org/doi/10.1103/PhysRevB.89.195134},
	doi = {10.1103/PhysRevB.89.195134},
	number = {19},
	urldate = {2016-11-08},
	journal = {Physical Review B},
	author = {Ferretti, Andrea and Dabo, Ismaila and Cococcioni, Matteo and Marzari, Nicola},
	month = may,
	year = {2014},
	pages = {195134}
}

@article{perdew_self-interaction_1981,
	title = {Self-interaction correction to density-functional approximations for many-electron systems},
	volume = {23},
	url = {http://link.aps.org/doi/10.1103/PhysRevB.23.5048},
	doi = {10.1103/PhysRevB.23.5048},
	number = {10},
	urldate = {2014-06-23},
	journal = {Physical Review B},
	author = {Perdew, J. P. and Zunger, Alex},
	year = {1981},
	pages = {5048--5079},
}

@article{stengel_self-interaction_2008,
	title = {Self-interaction correction with {Wannier} functions},
	volume = {77},
	url = {https://link.aps.org/doi/10.1103/PhysRevB.77.155106},
	doi = {10.1103/PhysRevB.77.155106},
	number = {15},
	urldate = {2017-06-19},
	journal = {Physical Review B},
	author = {Stengel, Massimiliano and Spaldin, Nicola A.},
	month = apr,
	year = {2008},
	pages = {155106},
}

@article{gatti_exciton_2013,
	title = {Exciton dispersion from first principles},
	volume = {88},
	issn = {1098-0121, 1550-235X},
	url = {https://link.aps.org/doi/10.1103/PhysRevB.88.155113},
	doi = {10.1103/PhysRevB.88.155113},
	language = {en},
	number = {15},
	urldate = {2021-03-10},
	journal = {Phys. Rev. B},
	author = {Gatti, Matteo and Sottile, Francesco},
	month = oct,
	year = {2013},
	pages = {155113},
}

@article{onida_electronic_2002,
	title = {Electronic excitations: density-functional versus many-body {Green}’s-function approaches},
	volume = {74},
	issn = {0034-6861, 1539-0756},
	shorttitle = {Electronic excitations},
	url = {https://link.aps.org/doi/10.1103/RevModPhys.74.601},
	doi = {10.1103/RevModPhys.74.601},
	language = {en},
	number = {2},
	urldate = {2020-03-19},
	journal = {Rev. Mod. Phys.},
	author = {Onida, Giovanni and Reining, Lucia and Rubio, Angel},
	month = jun,
	year = {2002},
	pages = {601--659}
}

@article{C3N_bonacci,
  title = {Excitonic effects in graphene-like ${\mathrm{C}}_{3}\mathrm{N}$},
  author = {Bonacci, Miki and Zanfrognini, Matteo and Molinari, Elisa and Ruini, Alice and Caldas, Marilia J. and Ferretti, Andrea and Varsano, Daniele},
  journal = {Phys. Rev. Mater.},
  volume = {6},
  issue = {3},
  pages = {034009},
  numpages = {7},
  year = {2022},
  month = {Mar},
  publisher = {American Physical Society},
  doi = {10.1103/PhysRevMaterials.6.034009},
  url = {https://link.aps.org/doi/10.1103/PhysRevMaterials.6.034009}
}

@article{Hedin_1965,
  title = {{New Method for Calculating the One-Particle Green's Function with Application to the Electron-Gas Problem}},
  author = {Hedin, Lars},
  journal = {Phys. Rev.},
  volume = {139},
  pages = {796--823},
  year = 1965,
  month = Aug,
  publisher = {American Physical Society},
  doi = {10.1103/PhysRev.139.A796},
}

@misc{k2y_github,
  author       = {Miki Bonacci},
  title        = {k2y code},
  year         = {2025},
  howpublished = {\url{https://github.com/mikibonacci/k2y}},
}

@article{GN-PPA,
  title = {Metal-insulator transition in Kohn-Sham theory and quasiparticle theory},
  author = {Godby, R. W. and Needs, R. J.},
  journal = {Phys. Rev. Lett.},
  volume = {62},
  issue = {10},
  pages = {1169--1172},
  numpages = {0},
  year = {1989},
  month = {Mar},
  publisher = {American Physical Society},
  doi = {10.1103/PhysRevLett.62.1169},
}

@article{HL-PPA,
  title = {Electron correlation in semiconductors and insulators: Band gaps and quasiparticle energies},
  author = {Hybertsen, Mark S. and Louie, Steven G.},
  journal = {Phys. Rev. B},
  volume = {34},
  issue = {8},
  pages = {5390--5413},
  numpages = {0},
  year = {1986},
  month = {Oct},
  publisher = {American Physical Society},
  doi = {10.1103/PhysRevB.34.5390},
}

@article{PIZZI2016,
title = {{AiiDA: automated interactive infrastructure and database for computational science}},
journal = {Computational Materials Science},
volume = {111},
pages = {218-230},
year = {2016},
issn = {0927-0256},

author = {Giovanni Pizzi and Andrea Cepellotti and Riccardo Sabatini and Nicola Marzari and Boris Kozinsky},

}

@article{huber2020aiida,
  title={{AiiDA 1.0, a scalable computational infrastructure for automated reproducible workflows and data provenance}},
  author={Huber, Sebastiaan P and Zoupanos, Spyros and Uhrin, Martin and Talirz, Leopold and Kahle, Leonid and H{\"a}uselmann, Rico and Gresch, Dominik and M{\"u}ller, Tiziano and Yakutovich, Aliaksandr V and Andersen, Casper W and others},
  journal={Scientific Data},
  volume={7},
  number={1},
  pages={300},
  year={2020},
  publisher={Nature Publishing Group UK London}
}

@article{UHRIN2021,
title = {{Workflows in AiiDA: Engineering a high-throughput, event-based engine for robust and modular computational workflows}},
journal = {Computational Materials Science},
volume = {187},
pages = {110086},
year = {2021},
issn = {0927-0256},
author = {Martin Uhrin and Sebastiaan P. Huber and Jusong Yu and Nicola Marzari and Giovanni Pizzi},

}

@misc{aiida_wannier90_workflows,
  title        = {aiida-wannier90-workflows},
  year         = {2025},
  howpublished = {\url{https://github.com/aiidateam/aiida-wannier90-workflows}},
}

@misc{aiida-quantumespresso,
  author       = {AiiDAteam},
  title        = {Official AiiDA plugin for Quantum ESPRESSO.},
  year         = {2025},
  howpublished = {\url{https://github.com/aiidateam/aiida-quantumespresso}},
}

@misc{aiida-koopmans,
  author       = {Miki Bonacci},
  title        = {aiida-koopmans},
  year         = {2025},
  howpublished = {\url{https://github.com/mikibonacci/aiida-koopmans}},
}

@article{baroni_DFPT,
  title = {Phonons and related crystal properties from density-functional perturbation theory},
  author = {Baroni, Stefano and de Gironcoli, Stefano and Dal Corso, Andrea and Giannozzi, Paolo},
  journal = {Rev. Mod. Phys.},
  volume = {73},
  issue = {2},
  pages = {515--562},
  numpages = {0},
  year = {2001},
  month = {Jul},
  publisher = {American Physical Society},
  doi = {10.1103/RevModPhys.73.515},
  url = {https://link.aps.org/doi/10.1103/RevModPhys.73.515}
}

@article{Kronik_optimally_tuned_2023,
  title = {Optical absorption spectra of metal oxides from time-dependent density functional theory and many-body perturbation theory based on optimally-tuned hybrid functionals},
  author = {Ohad, Guy and Gant, Stephen E. and Wing, Dahvyd and Haber, Jonah B. and Camarasa-G\'omez, Mar\'{\i}a and Sagredo, Francisca and Filip, Marina R. and Neaton, Jeffrey B. and Kronik, Leeor},
  journal = {Phys. Rev. Mater.},
  volume = {7},
  issue = {12},
  pages = {123803},
  numpages = {15},
  year = {2023},
  month = {Dec},
  publisher = {American Physical Society},
  doi = {10.1103/PhysRevMaterials.7.123803},
  url = {https://link.aps.org/doi/10.1103/PhysRevMaterials.7.123803}
}

@article{Schebek_2025,
doi = {10.1088/2515-7639/adab24},
url = {https://doi.org/10.1088/2515-7639/adab24},
year = {2025},
month = {feb},
publisher = {IOP Publishing},
volume = {8},
number = {2},
pages = {025006},
author = {Schebek, Maximilian and Pavone, Pasquale and Draxl, Claudia and Caruso, Fabio},
title = {Phonon-mediated renormalization of exciton energies and absorption spectra in polar semiconductors},
journal = {Journal of Physics: Materials},
}

@article{neaton_phonon_EBE_screening,
  title = {Phonon Screening of Excitons in Semiconductors: Halide Perovskites and Beyond},
  author = {Filip, Marina R. and Haber, Jonah B. and Neaton, Jeffrey B.},
  journal = {Phys. Rev. Lett.},
  volume = {127},
  issue = {6},
  pages = {067401},
  numpages = {7},
  year = {2021},
  month = {Aug},
  publisher = {American Physical Society},
  doi = {10.1103/PhysRevLett.127.067401},
  url = {https://link.aps.org/doi/10.1103/PhysRevLett.127.067401}
}

@article{Tal_tddft_2020,
  title = {Accurate optical spectra through time-dependent density functional theory based on screening-dependent hybrid functionals},
  author = {Tal, Alexey and Liu, Peitao and Kresse, Georg and Pasquarello, Alfredo},
  journal = {Phys. Rev. Res.},
  volume = {2},
  issue = {3},
  pages = {032019(R)},
  numpages = {6},
  year = {2020},
  month = {Jul},
  publisher = {American Physical Society},
  doi = {10.1103/PhysRevResearch.2.032019},
  url = {https://link.aps.org/doi/10.1103/PhysRevResearch.2.032019}
}

@article{Miglio_2020, title={Predominance of non-adiabatic effects in zero-point renormalization of the electronic band gap}, volume={6}, ISSN={2057-3960}, url={https://www.nature.com/articles/s41524-020-00434-z}, DOI={10.1038/s41524-020-00434-z}, number={1}, journal={npj Computational Materials}, author={Miglio, Anna and Brousseau-Couture, Véronique and Godbout, Emile and Antonius, Gabriel and Chan, Yang-Hao and Louie, Steven G. and Côté, Michel and Giantomassi, Matteo and Gonze, Xavier}, year={2020}, month=nov, pages={167}, language={en} }

@article{chen_2015,
  title = {Accurate band gaps of extended systems via efficient vertex corrections in $GW$},
  author = {Chen, Wei and Pasquarello, Alfredo},
  journal = {Phys. Rev. B},
  volume = {92},
  issue = {4},
  pages = {041115(R)},
  numpages = {5},
  year = {2015},
  month = {Jul},
  publisher = {American Physical Society},
  doi = {10.1103/PhysRevB.92.041115},
  url = {https://link.aps.org/doi/10.1103/PhysRevB.92.041115}
}

@article{tarrio_1989,
  title = {Interband transitions, plasmons, and dispersion in hexagonal boron nitride},
  author = {Tarrio, C. and Schnatterly, S. E.},
  journal = {Phys. Rev. B},
  volume = {40},
  issue = {11},
  pages = {7852--7859},
  numpages = {0},
  year = {1989},
  month = {Oct},
  publisher = {American Physical Society},
  doi = {10.1103/PhysRevB.40.7852},
  url = {https://link.aps.org/doi/10.1103/PhysRevB.40.7852}
}

@article{Thatribud_2019, title={Electronic and optical properties of TiO2 by first-principle calculation (DFT-GW and BSE)}, volume={6}, ISSN={2053-1591}, url={https://iopscience.iop.org/article/10.1088/2053-1591/ab2cad}, DOI={10.1088/2053-1591/ab2cad}, number={9}, journal={Materials Research Express}, author={Thatribud, A}, year={2019}, month=july, pages={095021} }

@article{shiskin_2007,
  title = {Accurate Quasiparticle Spectra from Self-Consistent GW Calculations with Vertex Corrections},
  author = {Shishkin, M. and Marsman, M. and Kresse, G.},
  journal = {Phys. Rev. Lett.},
  volume = {99},
  issue = {24},
  pages = {246403},
  numpages = {4},
  year = {2007},
  month = {Dec},
  publisher = {American Physical Society},
  doi = {10.1103/PhysRevLett.99.246403},
  url = {https://link.aps.org/doi/10.1103/PhysRevLett.99.246403}
}

@article{Soriano2018InterplayBI,
  title={Interplay between interlayer exchange and stacking in CrI3 bilayers},
  author={David Soriano and Claudia Cardoso and Joaqu{\'i}n Fern{\'a}ndez-Rossier},
  journal={Solid State Communications},
  year={2018},
  url={https://api.semanticscholar.org/CorpusID:53700797}
}

@article{Gibertini_2021,
doi = {10.1088/1361-6463/abc2f4},
url = {https://doi.org/10.1088/1361-6463/abc2f4},
year = {2020},
month = {nov},
publisher = {IOP Publishing},
volume = {54},
number = {6},
pages = {064002},
author = {Gibertini, Marco},
title = {Magnetism and stability of all primitive stacking patterns in bilayer chromium trihalides},
journal = {Journal of Physics D: Applied Physics}
}

@article{olsen_2021,
  title = {Unified Treatment of Magnons and Excitons in Monolayer ${\mathrm{CrI}}_{3}$ from Many-Body Perturbation Theory},
  author = {Olsen, Thomas},
  journal = {Phys. Rev. Lett.},
  volume = {127},
  issue = {16},
  pages = {166402},
  numpages = {6},
  year = {2021},
  month = {Oct},
  publisher = {American Physical Society},
  doi = {10.1103/PhysRevLett.127.166402},
  url = {https://link.aps.org/doi/10.1103/PhysRevLett.127.166402}
}

@article{Esquembre_2025,
  title = {Magnons in chromium trihalides calculated with the ab initio Bethe-Salpeter equation},
  author = {Esquembre-Ku\ifmmode \check{c}\else \v{c}\fi{}ukali\ifmmode \acute{c}\else \'{c}\fi{}, Ali and Le, Khoa B. and Garc\'{\i}a-Crist\'obal, Alberto and Bernardi, Marco and Sangalli, Davide and Molina-S\'anchez, Alejandro},
  journal = {Phys. Rev. B},
  volume = {112},
  issue = {18},
  pages = {184412},
  numpages = {13},
  year = {2025},
  month = {Nov},
  publisher = {American Physical Society},
  doi = {10.1103/gbpw-zh1v},
  url = {https://link.aps.org/doi/10.1103/gbpw-zh1v}
}

@article{strinati1988application,
  title = {Application of the Green’s functions method to the study of the optical properties of semiconductors},
  author = {Strinati, G},
  journal = {La Rivista del Nuovo Cimento (1978-1999)},
  volume = {11},
  number = {12},
  pages = {1--86},
  year = {1988},
  publisher = {Springer},
}

@article{Balapure_rev_photocat_chargeT_2024,
  author = {Balapure, Aniket and Ray Dutta, Jayati and Ganesan, Ramakrishnan},
  title = {Recent advances in semiconductor heterojunctions: a detailed review of the fundamentals of photocatalysis, charge transfer mechanism and materials},
  journal = {RSC Applied Interfaces},
  year = {2024},
  volume = {1},
  number = {1},
  pages = {43-69},
  month = {01},
  issn = {2755-3701},
  doi = {10.1039/d3lf00126a},
  url = {https://doi.org/10.1039/d3lf00126a},
  eprint = {https://pubs.rsc.org/lf/article-pdf/1/1/43/9150419/d3lf00126a.pdf},
}

@article{Yu_rev_plasmonics_2019,
  title = {Plasmon-enhanced light–matter interactions and applications},
  volume = {5},
  issn = {2057-3960},
  url = {https://www.nature.com/articles/s41524-019-0184-1},
  doi = {10.1038/s41524-019-0184-1},
  number = {1},
  journal = {npj Computational Materials},
  author = {Yu, Huakang and Peng, Yusi and Yang, Yong and Li, Zhi-Yuan},
  year = {2019},
  month = apr,
  pages = {45},
  language = {en},
}

@article{Wolf_exc_sensing_2026,
  title = {Exciton-based sensing of remote electron correlations in 2D heterostructures},
  author = {Wolf, Tobias M. R. and Xie, Tian and Jin, Chenhao and MacDonald, Allan H.},
  journal = {Phys. Rev. Lett.},
  pages = {},
  year = {2026},
  month = {Aug},
  publisher = {American Physical Society},
  doi = {10.1103/246f-2p98},
  url = {https://link.aps.org/doi/10.1103/246f-2p98},
}

@article{plasmon_exc_rev_Yi_2026,
  title = {Plasmon–exciton coupling in two-dimensional materials: fundamentals and applications in nanosensors},
  journal = {Sensor Review},
  volume = {46},
  number = {4},
  pages = {604-628},
  year = {2026},
  issn = {0260-2288},
  doi = {https://doi.org/10.1108/SR-08-2025-0591},
  url = {https://www.sciencedirect.com/science/article/pii/S0260228826000205},
  author = {Xiuyuan Yi},
}

@article{WANG_rev_exc_org_photocat_2026,
  title = {Exciton engineering in organic photocatalysis: mechanisms and applications},
  journal = {Environmental Functional Materials},
  volume = {5},
  number = {1},
  pages = {43-55},
  year = {2026},
  issn = {2773-0581},
  doi = {https://doi.org/10.1016/j.efmat.2025.12.002},
  url = {https://www.sciencedirect.com/science/article/pii/S2773058125000456},
  author = {Weiguang Wang and Shuai Gao and Jufang He and Jiahao Cui and Haodong Ji},
}

@article{Marzari_Ferretti_Wolverton_2021,
  title = {Electronic-structure methods for materials design},
  volume = {20},
  issn = {1476-1122, 1476-4660},
  url = {https://www.nature.com/articles/s41563-021-01013-3},
  doi = {10.1038/s41563-021-01013-3},
  number = {6},
  journal = {Nature Materials},
  author = {Marzari, Nicola and Ferretti, Andrea and Wolverton, Chris},
  year = {2021},
  month = june,
  pages = {736–749},
  language = {en},
}

@article{Solet_excitons_LiZnAs_ScAgC_photovoltaic_2025,
  title = {Many-body ab initio study of quasiparticles, optical excitations, and excitonic properties in $\mathrm{Li}\mathrm{Zn}\mathrm{As}$ and $\mathrm{Sc}\mathrm{Ag}\mathrm{C}$ for photovoltaic applications},
  author = {Solet, Vinod Kumar and Pandey, Sudhir K.},
  journal = {Phys. Rev. Appl.},
  volume = {23},
  issue = {6},
  pages = {064040},
  numpages = {18},
  year = {2025},
  month = {Jun},
  publisher = {American Physical Society},
  doi = {10.1103/z3dg-jxt1},
  url = {https://link.aps.org/doi/10.1103/z3dg-jxt1},
}

@article{Sharifzadeh_2018,
  title = {Many-body perturbation theory for understanding optical excitations in organic molecules and solids},
  volume = {30},
  issn = {0953-8984, 1361-648X},
  url = {https://iopscience.iop.org/article/10.1088/1361-648X/aab0d1},
  doi = {10.1088/1361-648X/aab0d1},
  number = {15},
  journal = {Journal of Physics: Condensed Matter},
  author = {Sharifzadeh, Sahar},
  year = {2018},
  month = apr,
  pages = {153002},
}

@article{Kumar_Singh_Gill_Bhattacharya_2021,
  title = {Optoelectronic Properties of Chalcogenide Perovskites by Many-Body Perturbation Theory},
  volume = {12},
  rights = {https://doi.org/10.15223/policy-029},
  issn = {1948-7185, 1948-7185},
  url = {https://pubs.acs.org/doi/10.1021/acs.jpclett.1c01034},
  doi = {10.1021/acs.jpclett.1c01034},
  number = {22},
  journal = {The Journal of Physical Chemistry Letters},
  author = {Kumar, Manish and Singh, Arunima and Gill, Deepika and Bhattacharya, Saswata},
  year = {2021},
  month = june,
  pages = {5301–5307},
  language = {en},
}

@article{KMICHOU2024112812,
  title = {Evaluation of the phonon, optoelectronic, photovoltaic, thermoelectric, photocatalytic and thermodynamic properties of the inorganic perovskite CsPbI2Br using Ab initio calculations},
  journal = {Solar Energy},
  volume = {279},
  pages = {112812},
  year = {2024},
  issn = {0038-092X},
  doi = {https://doi.org/10.1016/j.solener.2024.112812},
  url = {https://www.sciencedirect.com/science/article/pii/S0038092X24005073},
  author = {M. Kmichou and R. Masrour and L. Xu and E.K. Hlil and A. Rezzouk},
}

@article{setten_automationMBPT_2017,
  title = {Automation methodologies and large-scale validation for $GW$: Towards high-throughput $GW$ calculations},
  author = {van Setten, M. J. and Giantomassi, M. and Gonze, X. and Rignanese, G.-M. and Hautier, G.},
  journal = {Phys. Rev. B},
  volume = {96},
  issue = {15},
  pages = {155207},
  numpages = {11},
  year = {2017},
  month = {Oct},
  publisher = {American Physical Society},
  doi = {10.1103/PhysRevB.96.155207},
  url = {https://link.aps.org/doi/10.1103/PhysRevB.96.155207},
}

@article{Rasmussen_Deilmann_Thygesen_2021,
  title = {Towards fully automated GW band structure calculations: What we can learn from 60.000 self-energy evaluations},
  volume = {7},
  issn = {2057-3960},
  url = {https://www.nature.com/articles/s41524-020-00480-7},
  doi = {10.1038/s41524-020-00480-7},
  number = {1},
  journal = {npj Computational Materials},
  author = {Rasmussen, Asbjørn and Deilmann, Thorsten and Thygesen, Kristian S.},
  year = {2021},
  month = jan,
  pages = {22},
  language = {en},
}

@article{grosmann_robust_2024,
  title = {A robust, simple, and efficient convergence workflow for {GW} calculations},
  volume = {10},
  issn = {2057-3960},
  url = {https://www.nature.com/articles/s41524-024-01311-9},
  doi = {10.1038/s41524-024-01311-9},
  language = {en},
  number = {1},
  urldate = {2025-11-25},
  journal = {npj Computational Materials},
  author = {Großmann, Max and Grunert, Malte and Runge, Erich},
  month = jun,
  year = {2024},
  pages = {135},
}

@article{Grunert_Grossmann_Runge_2025,
  title = {Machine learning climbs the Jacob’s Ladder of optoelectronic properties},
  volume = {16},
  rights = {2025 The Author(s)},
  issn = {2041-1723},
  url = {https://www.nature.com/articles/s41467-025-63355-9},
  doi = {10.1038/s41467-025-63355-9},
  number = {1},
  journal = {Nature Communications},
  publisher = {Nature Publishing Group},
  author = {Grunert, Malte and Großmann, Max and Runge, Erich},
  year = {2025},
  month = aug,
  pages = {8142},
  language = {en},
}

@article{qiao_projectability_2023,
	title = {Projectability disentanglement for accurate and automated electronic-structure {Hamiltonians}},
	volume = {9},
	copyright = {2023 The Author(s)},
	issn = {2057-3960},
	url = {https://www.nature.com/articles/s41524-023-01146-w},
	doi = {10.1038/s41524-023-01146-w},
	language = {en},
	number = {1},
	urldate = {2026-09-01},
	journal = {npj Computational Materials},
	publisher = {Nature Publishing Group},
	author = {Qiao, Junfeng and Pizzi, Giovanni and Marzari, Nicola},
	month = nov,
	year = {2023},
	pages = {208},
}

@article{qiao_automated_2023,
	title = {Automated mixing of maximally localized {Wannier} functions into target manifolds},
	volume = {9},
	copyright = {2023 The Author(s)},
	issn = {2057-3960},
	url = {https://www.nature.com/articles/s41524-023-01147-9},
	doi = {10.1038/s41524-023-01147-9},
	language = {en},
	number = {1},
	urldate = {2026-09-01},
	journal = {npj Computational Materials},
	publisher = {Nature Publishing Group},
	author = {Qiao, Junfeng and Pizzi, Giovanni and Marzari, Nicola},
	month = oct,
	year = {2023},
	pages = {206},
}

@article{Ma_Wang_2016,
  title = {Using Wannier functions to improve solid band gap predictions in density functional theory},
  volume = {6},
  issn = {2045-2322},
  url = {https://www.nature.com/articles/srep24924},
  doi = {10.1038/srep24924},
  number = {1},
  journal = {Scientific Reports},
  author = {Ma, Jie and Wang, Lin-Wang},
  year = {2016},
  month = apr,
  pages = {24924},
  language = {en},
}

@article{Shen_ZPR_hBN_2020,
  author = {Shen, Tong and Zhang, Xiao-Wei and Shang, Honghui and Zhang, Min-Ye and Wang, Xinqiang and Wang, En-Ge and Jiang, Hong and Li, Xin-Zheng},
  title = {Influence of high-energy local orbitals and electron-phonon interactions on the band gaps and optical absorption spectra of hexagonal boron nitride},
  journal = {Phys. Rev. B},
  year = {2020},
  volume = {102},
  issue = {4},
  pages = {045117},
  numpages = {9},
  month = {Jul},
  publisher = {American Physical Society},
  doi = {10.1103/PhysRevB.102.045117},
  url = {https://link.aps.org/doi/10.1103/PhysRevB.102.045117},
}

@article{Cassabois_exp_hBN_2016,
  title = {Hexagonal boron nitride is an indirect bandgap semiconductor},
  volume = {10},
  issn = {1749-4885, 1749-4893},
  url = {https://www.nature.com/articles/nphoton.2015.277},
  doi = {10.1038/nphoton.2015.277},
  number = {4},
  journal = {Nature Photonics},
  author = {Cassabois, G. and Valvin, P. and Gil, B.},
  year = {2016},
  month = apr,
  pages = {262–266},
  language = {en},
}

@article{Gao_BSEDynPPA_MoS2_2016,
  title = {Dynamical Excitonic Effects in Doped Two-Dimensional Semiconductors},
  volume = {16},
  issn = {1530-6984, 1530-6992},
  url = {https://pubs.acs.org/doi/10.1021/acs.nanolett.6b02118},
  doi = {10.1021/acs.nanolett.6b02118},
  number = {9},
  journal = {Nano Letters},
  author = {Gao, Shiyuan and Liang, Yufeng and Spataru, Catalin D. and Yang, Li},
  year = {2016},
  month = sept,
  pages = {5568–5573},
  language = {en},
}

@dataset{bonacci_2026_hwmbd-5t734,
  author = {Bonacci, Miki and Linscott, Edward and Geiger, Julian and Marzari, Nicola and Colonna, Nicola},
  title = {Koopmans spectral functionals for excitons in materials },
  month = sep,
  year = {2026},
  publisher = {Materials Cloud},
  doi = {10.24435/materialscloud:jp-07},
  url = {https://doi.org/10.24435/materialscloud:jp-07},
}

\end{document}


\title{Supplementary Material: Koopmans spectral functionals for excitons in materials}
\author{Miki Bonacci}
\email[corresponding author: ]{miki.bonacci@nano.cnr.it}
\affiliation{Istituto Nanoscienze -- CNR, S3, Via G. Campi 213/A, Modena, Italy}
\affiliation{PSI Center for Scientific Computing, Theory and Data, 5232 Villigen PSI, Switzerland}
\affiliation{
National Centre for Computational Design and Discovery of Novel Materials (MARVEL), 5232 Villigen PSI, Switzerland
}
\author{Edward Linscott}
\affiliation{PSI Center for Scientific Computing, Theory and Data, 5232 Villigen PSI, Switzerland}
\affiliation{
National Centre for Computational Design and Discovery of Novel Materials (MARVEL), 5232 Villigen PSI, Switzerland
}
\author{Julian Geiger}
\affiliation{PSI Center for Scientific Computing, Theory and Data, 5232 Villigen PSI, Switzerland}
\affiliation{
National Centre for Computational Design and Discovery of Novel Materials (MARVEL), 5232 Villigen PSI, Switzerland
}
\author{Nicola Marzari}
\affiliation{PSI Center for Scientific Computing, Theory and Data, 5232 Villigen PSI, Switzerland}
\affiliation{
National Centre for Computational Design and Discovery of Novel Materials (MARVEL), 5232 Villigen PSI, Switzerland
}
\affiliation{Theory and Simulation of Materials (THEOS), \'Ecole Polytechnique F\'ed\'erale de Lausanne, CH-1015 Lausanne, Switzerland}
\affiliation{Theory of Condensed Matter, Cavendish Laboratory, University of Cambridge, Cambridge CB3 0US, United Kingdom}
\author{Nicola Colonna}
\affiliation{PSI Center for Scientific Computing, Theory and Data, 5232 Villigen PSI, Switzerland}
\affiliation{
National Centre for Computational Design and Discovery of Novel Materials (MARVEL), 5232 Villigen PSI, Switzerland
}
\maketitle
\glsresetall
\section{Convergence parameters}
Here we present the convergence parameters used for the \gls{bse} calculations. These are obtained via the \texttt{aiida-yambo}~\cite{bonacci2023} built-in workflows. In particular, the cutoffs for the summation over empty states $N_b$ and plane waves $G_{\rm cut}$ are taken from Ref.~\citenum{bonacci2023} (except for CrI$_3$, fully converged here), while the \textbf{k}-meshes are computed within this work. The calculations are converged by imposing a relative threshold of $\Delta\epsilon=10\%$ on the lowest excitonic eigenvalue $E^1$. For this step, the quasiparticle corrections are simulated by applying a scissor-and-stretching operator to the \gls{ks} eigenvalues, so as to reproduce the value of the band gap. This approximation is sufficient for the convergence tests. 
\begin{table}[htbp]
\centering
\renewcommand{\arraystretch}{1.2}
\setlength{\tabcolsep}{6pt}

\begin{tabular}{l|cc|c}

Material & N$_b$ & G$_{cut}$ & \textbf{k}-mesh \\

\cline{1-4}

\textbf{Silicon} & 400 & 16 & 12 x 12 x 12 \\

\textbf{ZnO} & 800 & 28 & 12 x 12 x 6 \\

\textbf{hBN} & 800 & 20 & 12 x 12 x 4 \\

\textbf{Diamond} & 300 & 20 & 8 x 8 x 8 \\

\textbf{TiO$_2$} & 600 & 12 & 8 x 8 x 12 \\

\textbf{CrI$_3$} & 800 & 28 & 4 x 4 x 4 \\

\end{tabular}

\caption{Convergence parameters for the Yambo calculations.} 
\label{tab:SI_params}
\end{table}
\section{Relation between the band gaps and the excitonic peaks}
\begin{table}[htbp]
\centering
\renewcommand{\arraystretch}{1.2}
\setlength{\tabcolsep}{6pt}

\begin{tabular}{l|cc|cc|cc|cc}
& \multicolumn{2}{c|}{$E_g^{\mathrm{min}}$ (eV)} 
& \multicolumn{2}{c|}{$E_g^{\mathrm{dir}}$ (eV)} 
& \multicolumn{2}{c|}{$E^1$ (eV)} 
& \multicolumn{2}{c}{\glsentryshort{ebe} (eV)} \\
\cline{1-9}
Material 
& qKI & G$_0$W$_0$ 
& qKI & G$_0$W$_0$ 
& qKI & G$_0$W$_0$
& qKI & G$_0$W$_0$ \\
\midrule

\textbf{Silicon} 
& 1.32 & 1.17 
& 3.44 & 3.25   
& 3.55 & 3.35
& 0.08 & 0.09 \\

\textbf{ZnO}   
& 3.6  & 2.32   
& 3.6  & 2.32 
& 3.43 & 2.16
& 0.19 & 0.20 \\

\midrule

\textbf{hBN}     
& 7.10 & 6.15 
& 7.61 & 6.46 
& 6.78 & 5.51
& 1.0 & 1.11 \\

\textbf{Diamond}   
& 7.0  & 5.58 
& 8.32   & 7.28
& 8.13 & 6.98
& 0.35 & 0.30 \\

\textbf{TiO$_2$}
& 4.07  & 3.21   
& 4.07 & 3.25 
& 4.73 & 3.91
& 0.10 & 0.11 \\

\bottomrule
\end{tabular}

\caption{Comparison of \gls{qki} and $G_0W_0$ results for the minimum (fundamental) band gap $E_g^{\mathrm{min}}$, the minimum direct band gap $E_g^{\mathrm{dir}}$, the energy of the lowest excitonic peak $E^1$, and the corresponding \gls{ebe}. The \gls{ebe} is defined with respect to the independent-particle electron--hole transition associated with the excitonic state and therefore does not necessarily coincide with $E_g^{\mathrm{dir}}-E^1$. The slightly indirect nature of the $G_0W_0$ band gap of TiO$_2$ has been previously reported~\cite{Louie_TiO2}.}
\label{tab:SI_results}
\end{table}
In \cref{tab:SI_results}, we compare the minimum (fundamental) band gap $E_g^{\mathrm{min}}$, the minimum direct band gap $E_g^{\mathrm{dir}}$, the energy of the lowest excitonic peak $E^1$, and the \gls{ebe} associated with this excitonic state. Here, the \gls{ebe} is defined as the energy difference between the independent-particle electron-hole transition associated with the excitonic state and the corresponding interacting \gls{bse} eigenvalue, as described in the main text. It therefore does not necessarily coincide with the difference between the minimum direct gap $E_g^{\mathrm{dir}}$ and $E^1$, since the lowest excitonic state may involve transitions away from the point in the \gls{bz} at which the minimum direct gap occurs.
Nevertheless, a clear correspondence emerges between the differences in the direct quasiparticle gaps and those in the excitonic peak energies. Since the \gls{bse} describes vertical electron-hole transitions, the quasiparticle transition energies entering its diagonal part correspond to direct transitions, while the electron-hole interaction gives rise to the excitonic renormalization. The \gls{ebe} can therefore be viewed, to a good approximation, as the red shift of the relevant direct quasiparticle transition associated with the excitonic state. The close agreement between the \gls{ebe} obtained with qKI and $G_0W_0$ consequently indicates that the differences in $E^1$ primarily reflect differences in the corresponding direct quasiparticle transition energies.
The hBN case provides a particularly clear illustration of this behavior. The minimum direct gaps obtained with qKI and $G_0W_0$ differ by 1.15~eV, while the corresponding first excitonic peaks differ by 1.27~eV. In contrast, the associated \glspl{ebe} differ by only 0.11~eV. The substantial discrepancy in $E^1$ is therefore due to the different direct quasiparticle transition energies entering the \gls{bse}, rather than to a significantly different electron-hole interaction. A similar trend is observed for diamond. These results further support the picture that, within the level of convergence adopted here, the difference between the BSE@qKI and BSE@$G_0W_0$ excitonic energies is predominantly inherited from the different quasiparticle transition energies, while the electron-hole interaction produces a similar renormalization in the two approaches. The origin of the remaining discrepancy between the qKI and $G_0W_0$ direct gaps, particularly for diamond, requires further investigation and is left for future work.

It is worth noting that the convergence requirements for the \gls{ebe} can be more stringent than those for the excitonic peak energy, particularly for systems with small binding energies. A more systematic and tighter convergence of the electron-hole transition space would therefore be required for a quantitative assessment of the \gls{ebe}, but this is beyond the scope of the present work. Our aim here is instead to establish that BSE@qKI can reproduce the excitonic properties obtained with the standard BSE@$G_0W_0$ approach within the accuracy and level of convergence adopted in this study.

\section{Absorption spectra and lowest exciton composition for all the materials studied in this work}

In the following we show the absorption spectra (the imaginary part of the macroscopic dielectric function $\epsilon_M$) obtained from the \gls{bse}, starting from the band structures computed via qKI and $G_0W_0$. The analogous \gls{rpa} spectrum (\emph{i.e.}, without the electron-hole interaction kernel) is also provided for the qKI case, together with experimental spectra as taken from the literature (see Table~1 of the main text).

\def\materials{
    Si2/Silicon,
    O2Zn2/ZnO,
    B2N2/hBN,
    C2/Diamond,
    TiO2/TiO$_2$
}

\foreach \material/\alias in \materials{
\begin{figure}[htbp]
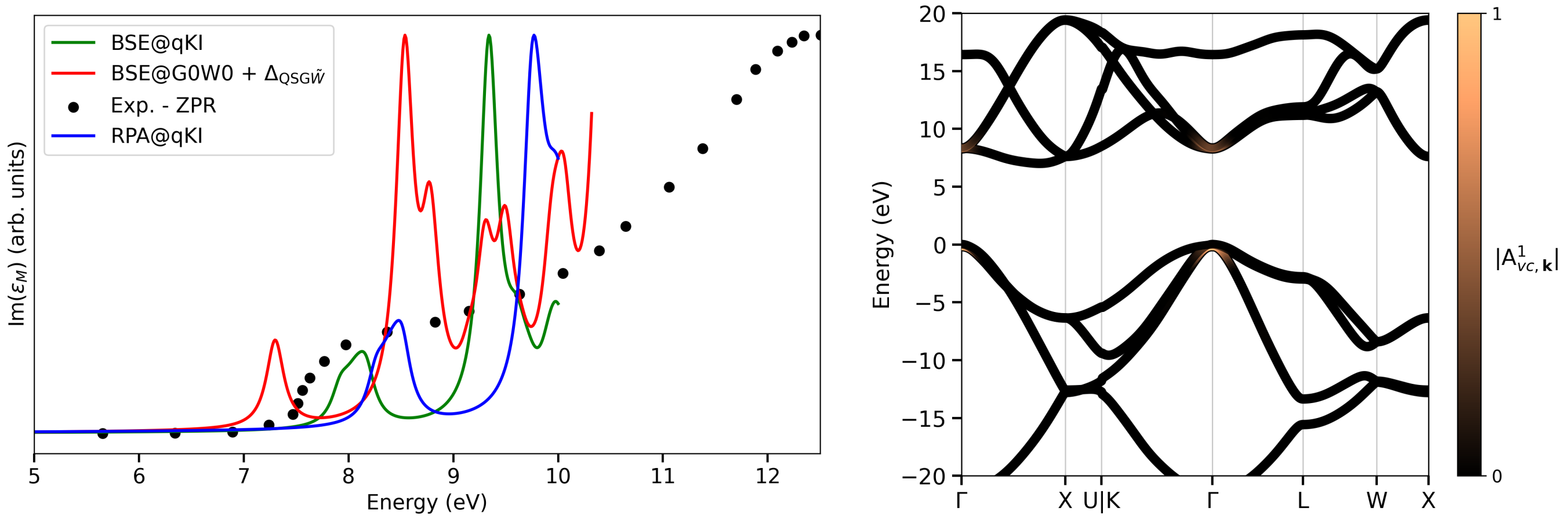

    \centering

    \begin{subfigure}{0.53\textwidth}
        \centering
        \includegraphics[width=\textwidth]{figures/\material_BSE_KI_GW_shifted_noZPR.png}
    \end{subfigure}
    \hfill
    \begin{subfigure}{0.45\textwidth}
        \centering
        \includegraphics[width=\textwidth]{figures/\material_exciton_fatbands_cmap.png}
    \end{subfigure}

    \caption{Absorption spectrum (left panel) and composition of the first Im($\epsilon_M$) peak, E$^1$ (right panel), for \alias, plotted on top of the qKI interpolated band structure. The experimental spectrum is shifted by -\glsentryshort{zpr} to be compared with theoretical results.}
    \label{fig:double_\material}
\end{figure}
}
\begin{figure}[htbp]
    \centering

    \begin{subfigure}{0.53\textwidth}
        \centering
        \includegraphics[width=\textwidth]{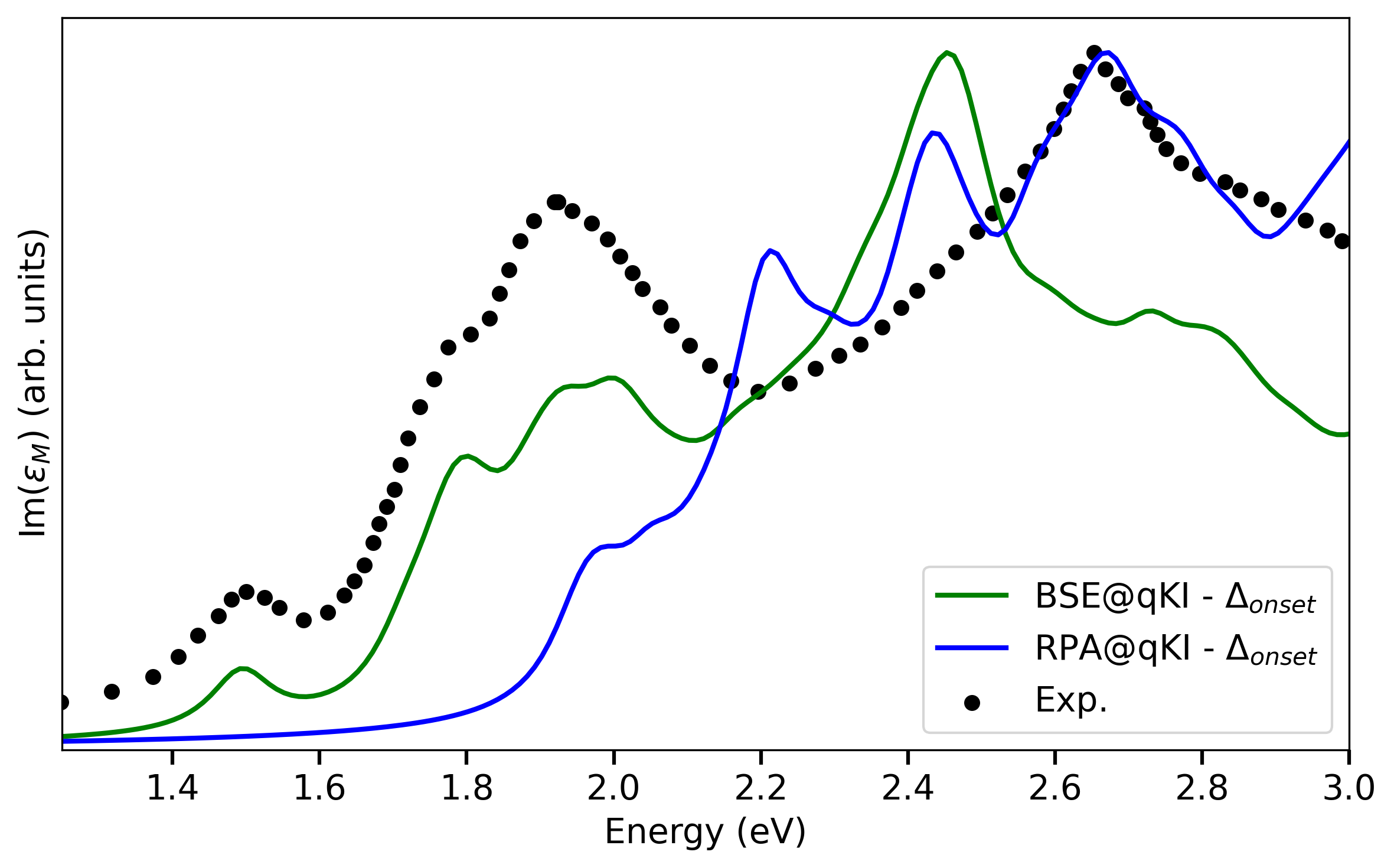}
    \end{subfigure}
    \hfill
    \begin{subfigure}{0.45\textwidth}
        \centering
        \includegraphics[width=\textwidth]{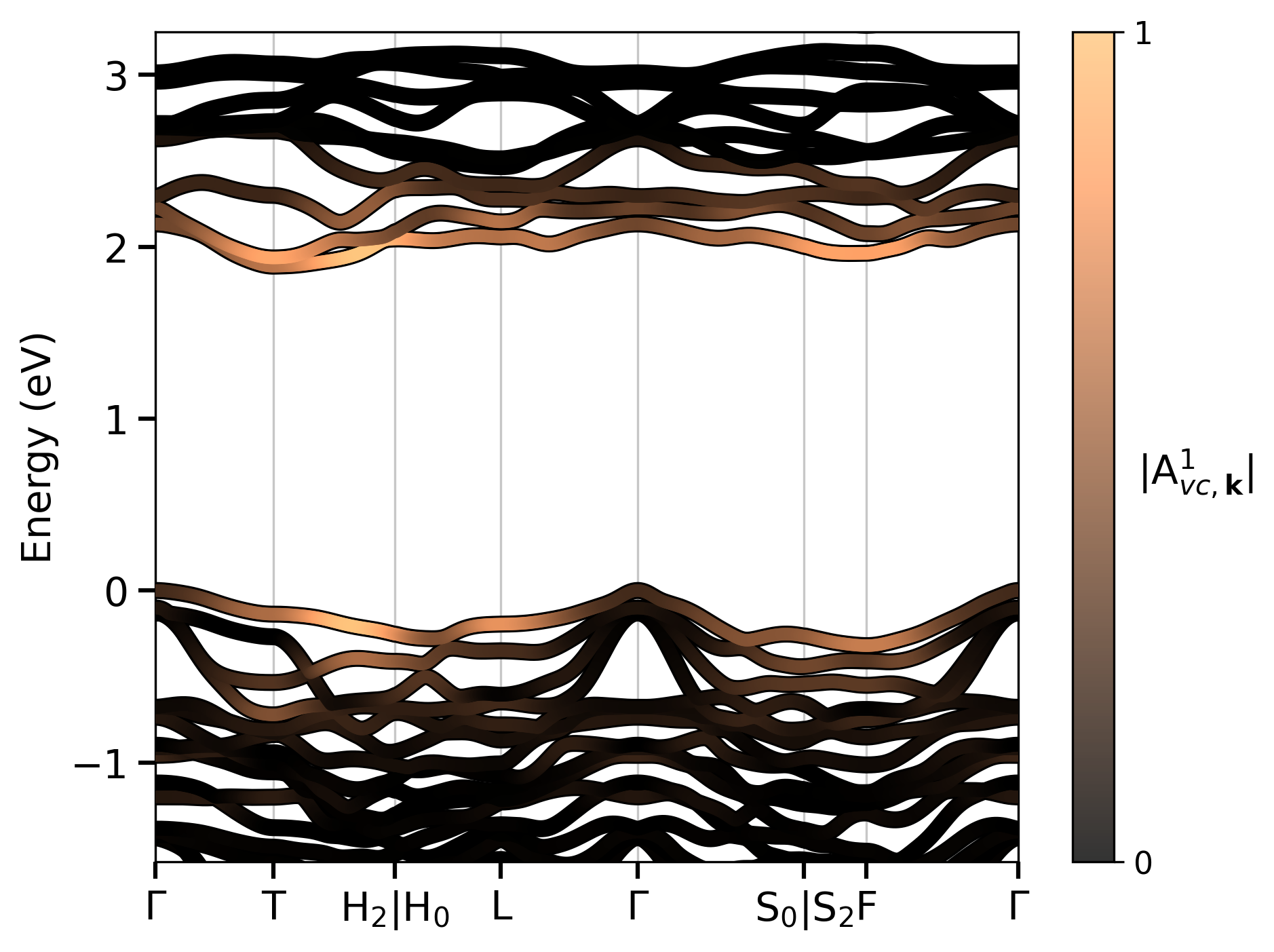}
    \end{subfigure}

    \caption{Absorption spectrum and composition of the first Im($\epsilon_M$) peak, E$^1$, for CrI$_3$, plotted on top of the qKI interpolated band structure. Here, at variance with the other cases, the theoretical spectra are shifted by $\Delta_{onset}$ to align with the experimental measurement.}
    \label{fig:SI_CrI3_spectrum}
\end{figure}
\section{Wannier-interpolated band structure for all the materials studied in this work}
Here we present the results for the interpolated band structures. The interpolation is done via the \texttt{aiida-yambo-wannier90}~\cite{bonacci2023} plugin. We note, in particular, the inaccuracy of the interpolation for the $G_0W_0$ band structure of ZnO, which is mostly due to the well-known inaccuracy of standard $G_0W_0$ in this particular case~\cite{Dario_ZnO_2024}. The BSE@$G_0W_0$ results are not affected by this, however, as they are mainly concentrated on the band structure close to the Fermi energy around the $\Gamma$ point.
\def\materials{
    Si2/Silicon ,
    O2Zn2/ZnO ,
    B2N2/hBN ,
    C2/Diamond ,
    TiO2/TiO$_2$ 
}

\foreach \material/\alias in \materials{
\begin{figure}[htbp]
    \centering
    \includegraphics[width=0.45\textwidth]{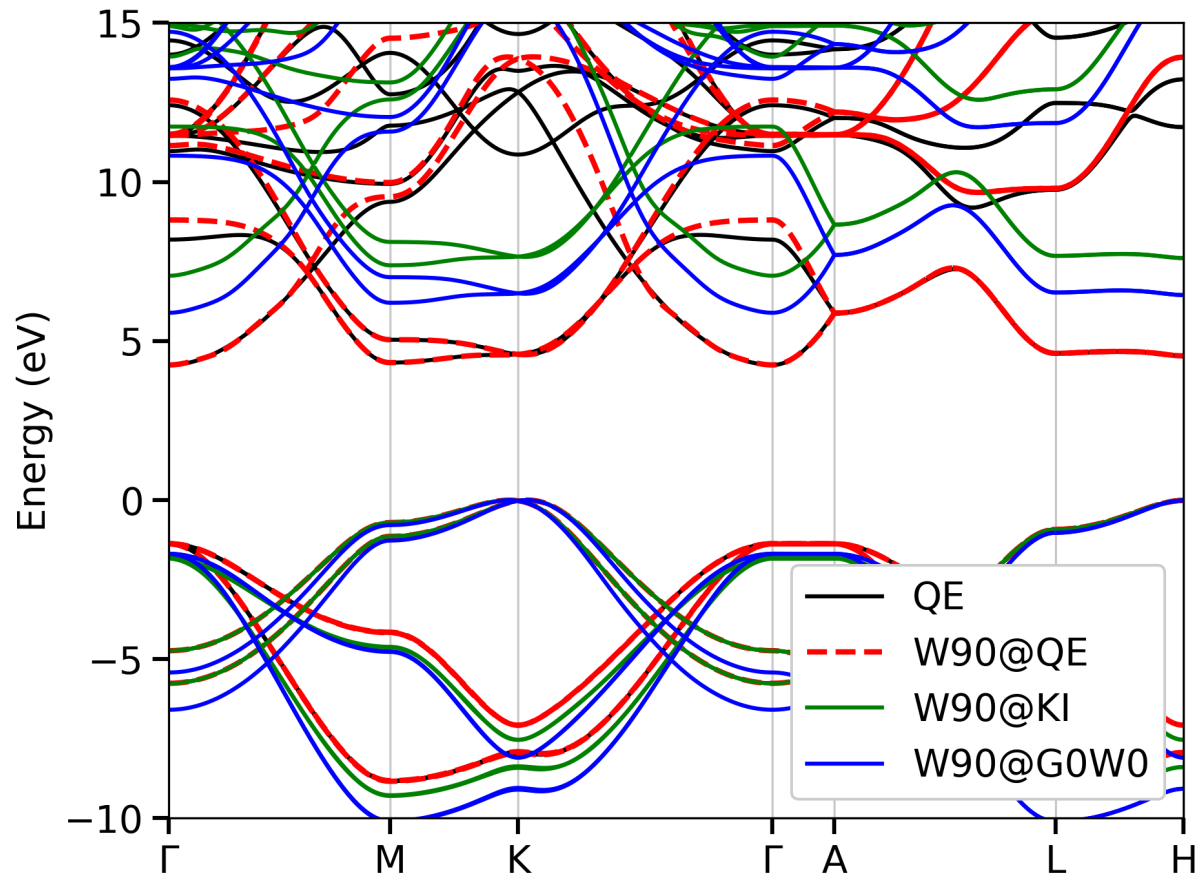}
    \caption{Interpolated band structure of \alias using different methods: \glsentryshort{dft} (QE), qKI Koopmans functionals (KI) and $G_0W_0$. The plot also shows the \QE\ bands computed on the same path.}
    \label{fig:wannier_interpolated_\material}
\end{figure}
}
\begin{figure}[htbp]
    \centering
    \includegraphics[width=0.5\textwidth]{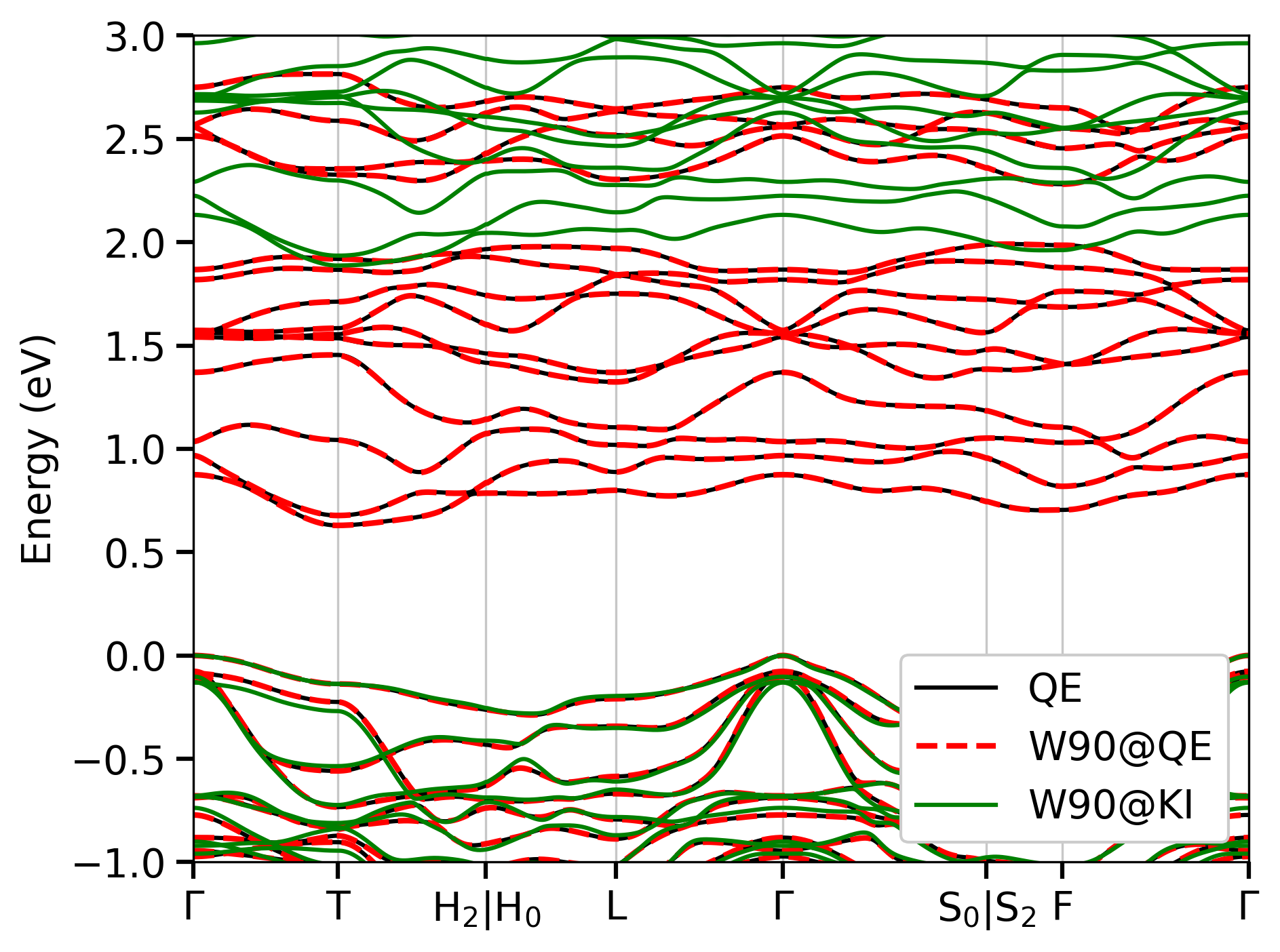}
    \caption{Interpolated band structure of CrI$_3$ using the qKI method. The plot also shows the \QE\ bands computed on the same path.}
    \label{fig:SI_CrI3_bands}
\end{figure}
\bibliographystyle{my_aip}
\bibliography{bib}